\documentclass[10pt,draftcls,onecolumn]{IEEEtran}
\usepackage{cite}
\ifCLASSINFOpdf
  \usepackage[pdftex]{graphicx}
  \DeclareGraphicsExtensions{.pdf,.jpeg,.png}
\else
\fi
\usepackage[cmex10]{amsmath}
\usepackage{amsfonts,amssymb,amsthm,mdwmath,bbm}
\newtheorem{remark}{Remark}
\newtheorem{lemma}{Lemma}
\usepackage{array}
\ifCLASSOPTIONcompsoc
  \usepackage[caption=false,font=normalsize,labelfont=sf,textfont=sf]{subfig}
\else
  \usepackage[caption=false,font=footnotesize]{subfig}
\fi
\usepackage{stfloats}
\usepackage{url}
\usepackage{color}
\usepackage{acronym}

\begin{document}
\acrodef{PVT}[PVT]{Poisson Voronoi Tessellation}
\acrodef{dof}[dof]{degrees of freedom}
\acrodef{PPP}[PPP]{Poisson Point Process}
\acrodef{CDF}[CDF]{Cumulative Distribution Function}
\acrodef{PDF}[PDF]{Probability Distribution Function}
\acrodef{PMF}[PMF]{Probability Mass Function}
\acrodef{WSN}[WSN]{Wireless Sensor Networks}
\acrodef{RV}[RV]{Random Variable}
\acrodef{AP}[AP]{Access point}

\title{Distribution of Cell Area  in Bounded Poisson Voronoi Tessellations with Application to Secure Local Connectivity}
\author{Konstantinos Koufos and Carl P. Dettmann 
\thanks{K.~Koufos and C.P.~Dettmann are with the School of Mathematics, University of Bristol, BS8 1TW, Bristol, UK. \{K.Koufos, Carl.Dettmann\}@bristol.ac.uk}} 
\maketitle

\begin{abstract}
Poisson Voronoi tessellations have been used in modeling many types of systems across different sciences, from geography and astronomy to telecommunications. The existing literature on the statistical properties of Poisson Voronoi cells is vast, however, little is known about the properties of Voronoi cells located close to the boundaries of a compact domain. In a domain with boundaries, some Voronoi cells would be naturally clipped by the boundary, and the cell area falling inside the deployment domain would have different statistical properties as compared to those of non-clipped Voronoi cells located in the bulk of the domain. In this paper, we consider the planar Voronoi tessellation induced by a homogeneous Poisson point process of intensity $\lambda\!>\!0$ in a quadrant, where the two half-axes represent boundaries. We show that the mean cell area is less than $\lambda^{-1}$ when the seed is located exactly at the boundary, and it can be larger than $\lambda^{-1}$ when the seed lies close to the boundary. In  addition, we calculate the second moment of cell area at two locations for the seed: (i) at the corner of a quadrant, and (ii) at the boundary of the half-plane. We illustrate that the two-parameter Gamma distribution, with location-dependent parameters calculated using the method of moments, can be of use in approximating the distribution of cell area. As a potential application, we use the Gamma approximations to study the degree distribution for secure connectivity in wireless sensor networks deployed over a domain with boundaries. 
\end{abstract}
\begin{IEEEkeywords}
Clipped Voronoi cells, physical layer security, Poisson Voronoi tessellations, stochastic geometry.
\end{IEEEkeywords}

\section{Introduction} 
A random tessellation is a random subdivision of a space into disjoint regions or cells $\mathcal{C}_i$, see~\cite{Moller1989, Lieshout2012} for a formal definition. Perhaps the most basic random tessellation model partitions the plane $\mathbb{R}^2$ into Voronoi cells. In order to construct them, a set of random nuclei (or seeds) $S_i$ are first distributed, and then, the locations of the plane are associated with the nearest seed for the Euclidean distance. The boundaries of the Voronoi cells are equidistant to the two nearest seeds, while the vertices of the tessellation are equidistant to the three nearest seeds. When the seeds are distributed randomly with intensity $\lambda\!>\!0$, i.e., a \ac{PPP}, the random tessellation is widely-known as the \ac{PVT}~\cite{Moller1989, Lieshout2012}.  

Since the concept of Voronoi tessellations is quite fundamental, it accepts a wide range of applications across the sciences. In  seismology, the seeds may represent epicentral locations of earthquakes. The tapered Pareto distribution, which favours the extreme events less than the Pareto distribution, was found to model well the distribution of Voronoi cell areas~\cite{Schoenberg2008}. In biology, the seeds may represent nest sites. The area of influence (area of Voronoi cell) was found to carry important biological information, e.g., it can be used as a measure for the breeding success~\cite{Schlicht2014}. In astrophysics, the seeds may represent galaxies, and the area of a cell can be used to estimate the density of galaxies at that location. Adjacent cells with density higher than a threshold are grouped together, thereby clusters of galaxies can be separated from the background~\cite{Ramella2001}. Voronoi tessellations are also useful for data partition, visualization and analysis, because they carry more information in comparison with their binned data counterparts~\cite{Cappellari2009}. Recently, Voronoi binning has also been proposed for analyzing the outcome of high-energy particle colliders. Each outcome is represented by a point in the phase space, labelled with information about momentum of particles, etc. The point process governing the resulting tessellation is in general a non-homogeneous \ac{PPP}. The relative standard deviation of the area of neighboring Voronoi cells could be used as an indicator for identifying the edge cells, which separate regions of different intensities in the phase space. In this way, different outcomes can be categorized~\cite{Debnath2015}. 

The statistical properties of planar \acp{PVT}, e.g., cell area, perimeter, vertex degree, etc. have been studied since the early 1950's~\cite{Meijering1952, Gilbert1962}. The \ac{PDF} of the area of the typical (randomly selected) cell in a planar \ac{PVT} is unknown, and approximations using the Gamma and the log-normal distribution with appropriately selected parameters have been widely adopted~\cite{Weaire1986,Tanemura2003,Kumar1992,Szabo2007}. The quality of these approximations has been mostly established by simulations. An intuitive explanation for the good fit of the Gamma distribution, based on the distribution of nearest neighbors for a planar \ac{PPP}, is claimed in~\cite{Weaire1986}. In order to avoid heavy simulations, an integral-based method is devised in~\cite{Brakke1986}, which is used to compute various second-order statistics of the \ac{PVT} including the edge length, the \acp{PDF} of the distance and angle between neighboring seeds and vertices, the area of the typical cell, etc. Unlike the \ac{PVT}, the distribution of cell area in planar Poisson Delaunay tessellation (the dual graph of the Voronoi diagram) is known; it can be expressed in terms of the modified Bessel function~\cite{Rathie1992}. For three-dimensional Delaunay cells, some properties of geometrical characteristics are available in~\cite{Muche1996}. For the statistical properties of three-dimensional Voronoi cells, see~\cite{Lazar2013} and references therein. 

If we would like to partition a bounded domain into Poisson Voronoi cells, some of the cells would be naturally clipped by the boundaries, see for instance~\cite[Fig.~1]{Yan2013}. In practice, we would be mostly interested in the parts of the cells falling inside the domain, i.e., the intersection of the \ac{PVT} and the domain. Intuitively, the statistics, e.g., cell area, of the clipped cells which are located close to the boundaries, would be different in comparison with those located in the bulk (non-clipped cells). We are motivated to study the distribution of the area falling within the deployment domain for clipped Voronoi cells, because wireless communication networks are embedded in domains with boundaries. Since the Voronoi partition of any closed object gives rise to clipped cells, other potential applications are also evident.

In wireless communication networks, \acp{PVT} have been used to describe the coverage area of base stations and/or sensors deployed in an irregular structure. The coverage area calculated in this way ignores the impact of fading in the wireless channel, the interference and the transmit power level, which might be different at different locations. Nevertheless, the \ac{PVT} is widely accepted by network engineers, and it is often used to get input into more complicated models including interference, e.g., the shot noise~\cite{Baccelli2001}. The Gamma approximation for the distribution of cell area in planar \acp{PVT} without boundaries has been used so far in cellular systems studies, e.g., by converting cell area distribution to network load distribution for a typical (randomly selected) base station~\cite{Cao2013}, as well as in \ac{WSN}, e.g., to investigate the \ac{PDF} of the number of secure communication links towards an \ac{AP}~\cite{Pinto2012}; a more detailed explanation for secure connectivity would be presented in Section~\ref{sec:Secrecy}. These studies use the Gamma parameters suggested in~\cite{Szabo2007}, etc., which correspond to a \ac{PVT} in $\mathbb{R}^2$, or equivalently, to wireless networks with infinite extent. This assumption might not be representative for all communication scenarios and/or for all the base stations in the network. For instance, realizing that most of the wireless data traffic is consumed  indoors, we witness nowadays the deployment of low-power base stations inside buildings and shopping malls. Similarly, \acp{WSN} are deployed indoors to collect and communicate data measurements necessary for automated applications. In this kind of scenarios, it might not be possible to neglect boundary effects. 

In this paper, we consider a homogeneous \ac{PPP} $\mathcal{S}$ of finite intensity $\lambda$ over the quadrant $\mathbb{R}^2_+$, where the two half-axes represent physical boundaries. The points (or seeds) of the process are denoted by $S_i, i\!=\!1,2,\ldots$ We also assume an additional point $S_0$, which is arbitrarily fixed either at the boundary or close to that. Let us denote by $\mathcal{C}_0$ the Voronoi cell centered at $S_0$ with respect to the set $\mathcal{S}\cup \left\{S_0\right\}$. Firstly, we compute the mean area for the Voronoi cell $\mathcal{C}_0$ falling inside the quadrant. We show that it can be smaller or larger than $\lambda^{-1}$, depending on the location of the seed $S_0$. On the other hand, in $\mathbb{R}^2$, the mean area of Voronoi cells is $\lambda^{-1}$. Secondly, we extend the method of~\cite{Brakke1986} to compute the second moment of the area, after fixing $S_0$ either at the corner of the quadrant or at the boundary and far from the corner. The latter can be seen as a \ac{PVT} in the half-plane with the point $S_0$ located at the boundary. In both cases, we illustrate that the two-parameter Gamma distribution with fitted mean and variance using the method of moments can be a useful approximation for the distribution of the area of $\mathcal{C}_0$ falling inside the domain. The parameters of the Gamma approximation depend on the location of $S_0$. Even though the distribution of the area is not exactly Gamma, as we will see a discrepancy around the peak, the approximations can be incorporated into certain applications introducing negligible errors. For instance, we will use the approximations in \acp{WSN} to study the distribution of the number of sensors with secure connection to and from an \ac{AP} in the presence of eavesdroppers.

In a recent paper~\cite{Devroye2017}, it has been shown that the asymptotic distribution (as the number of points tending to infinity) of the Voronoi cell area is independent of the location of the seed $S_0$ (almost everywhere) and of the intensity measure underlying the \ac{PPP}, including also the case of inhomogeneous \ac{PPP}. Our results complement the analysis in~\cite{Devroye2017}, showing that for a \ac{PPP} with finite intensity $\lambda$, the moments of the cell area are location-dependent near the boundaries.

The rest of the paper is organized as following: In Section~\ref{sec:PVT}, we present the notation. In Section~\ref{sec:Mean}, we prove the main outcome of this paper, i.e., in clipped \acp{PVT}, the mean cell area falling inside the deployment domain is location-dependent. In Section~\ref{sec:Variance}, we show how to calculate the variance of cell area using numerical integration, and we identify the parameters for the gamma \ac{PDF} approximating the cell area distribution for seeds fixed at the boundary. In Section~\ref{sec:Secrecy}, we use the gamma approximation to study properties for the secrecy graph in \acp{WSN} deployed in domains with borders. In Section~\ref{sec:Conclusions}, we conclude this study.

\section{Poisson Voronoi tessellation over a quadrant}
\label{sec:PVT}
\begin{figure}
\centering
  \includegraphics[width=3in]{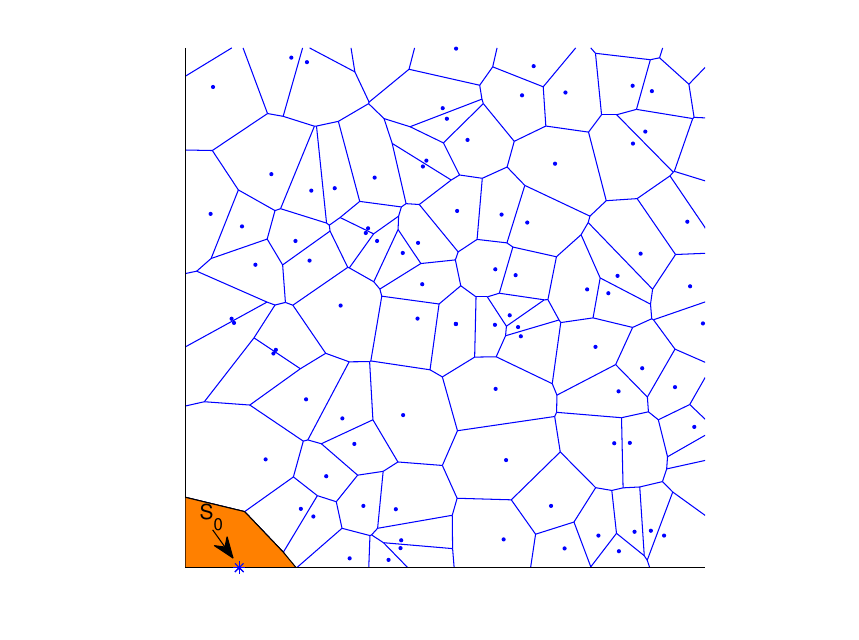}
 \caption{The dots represent a realization of a \ac{PPP} of unit intensity $\lambda\!=\!1$. Seed $S_0$ (asterisk) is added along the boundary at $\left(1,0\right)$, or $a\!=\!1$. The Voronoi cell $\mathcal{C}_0$ is clipped from the boundaries. The area of $\mathcal{C}_0$ inside the quadrant is colored.}
 \label{fig:S0Clipped}
\end{figure}

We consider a \ac{PPP} of unit intensity (without loss of generality) over the quadrant $\left[0,\infty \right)\times\left[0,\infty\right)$. We denote by $\mathcal{S}$ the set of points $S_i, i\!=\!1,2,\ldots$ generated by the process. In addition, we place a point $S_0$ at a fixed location, either (i) along the boundary at distance $a\!\geq\!0$ from the corner of the quadrant or, (ii) at distance $h\!\geq\!0$ from the boundary of the half-plane. The latter can also be seen as the case where $S_0$ is located far from the corner of the quadrant and at distance $h$ from the boundary. We consider the intersection of the Voronoi tessellation with respect to the set  $\mathcal{S}\cup\left\{S_0\right\}$ and the quadrant. Let $\mathcal{C}_0$ be the Voronoi cell of seed $S_0$. Due to the fact that $S_0$ is located at the boundary or close to that, the cell $\mathcal{C}_0$ would be probably clipped. We are interested in the distribution (over all realizations of the \ac{PPP} $\mathcal{S}$) of the area of $\mathcal{C}_0$ falling inside the quadrant, see Fig.~\ref{fig:S0Clipped} for a snapshot. Note that due to Slivnyak's Theorem~\cite{Stoyan1995}, the properties of $S_0$ in the process $\mathcal{S}\cup\left\{S_0\right\}$ are the same with those of  $S_i, i\!\in\!\left\{1,2,\ldots\right\}$ in the \ac{PPP} $\mathcal{S}$, conditioned on $S_i$ located at $S_0$. 

Let us consider a general point $P\in\mathbb{R}^2_+$ with polar coordinates $\left(r,\phi\right)$. The point $P$ can be interior to some Voronoi cell, at the boundary separating two cells, or it can also be a vertex. Adopting the terminology used in~\cite{Brakke1986}, we define the {\it{void}} of $P$ to be the intersection of the quadrant $\mathbb{R}^2_+$, and the disk $D$ with center $P$ and radius equal to the distance $d$ between $P$ and the nearest seed(s), $D\left(P,\min_i d\left(P,S_i\right)\right), i\!=\!0,1,2,\ldots$ The area of the void is denoted by $V\!\left(P\right)$. We denote by $A$ the area of the cell $\mathcal{C}_0$ falling inside the quadrant. In the next section, we show how to calculate the mean area $\mathbb{E}\left\{A\right\}$. 

\section{Mean cell area $\mathbb{E}\!\left\{A\right\}$}
\label{sec:Mean}
\begin{figure*}[!t]
\centering
\subfloat[]{\includegraphics[width=3.0in]{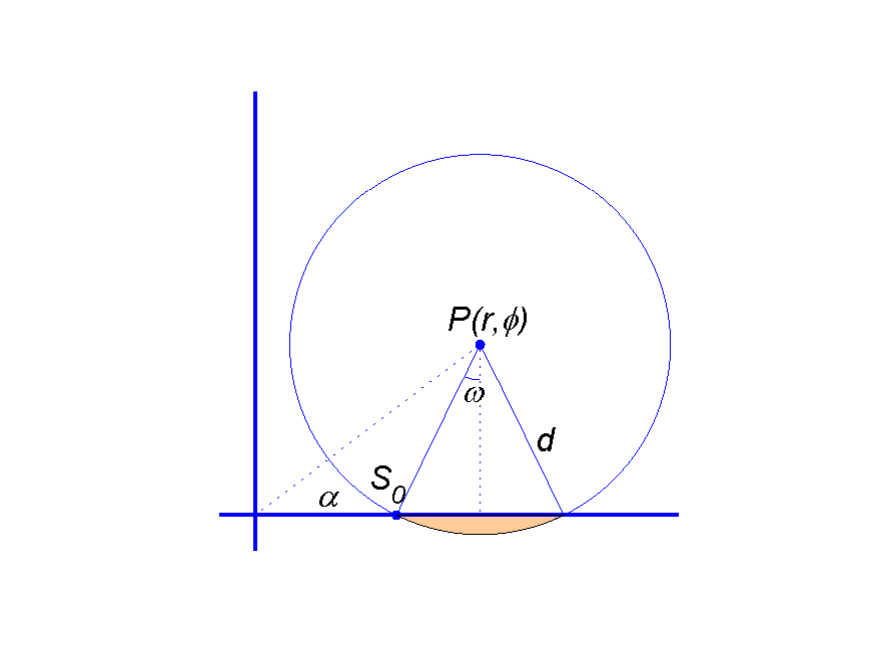}\label{fig:Integral1}} \hfil  
\subfloat[]{\includegraphics[width=3.0in]{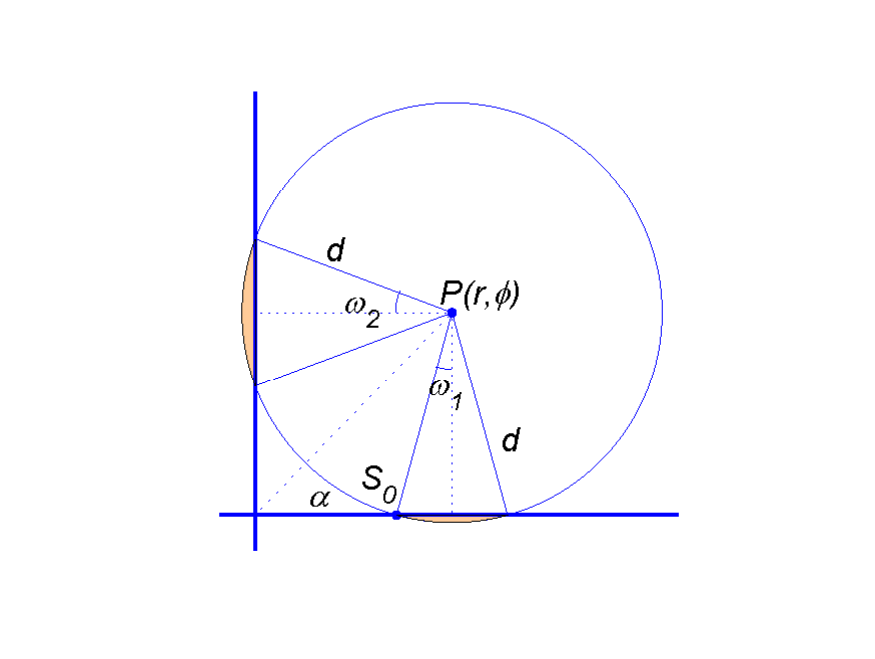}\label{fig:Integral2}} 
\\ \subfloat[]{\includegraphics[width=3.0in]{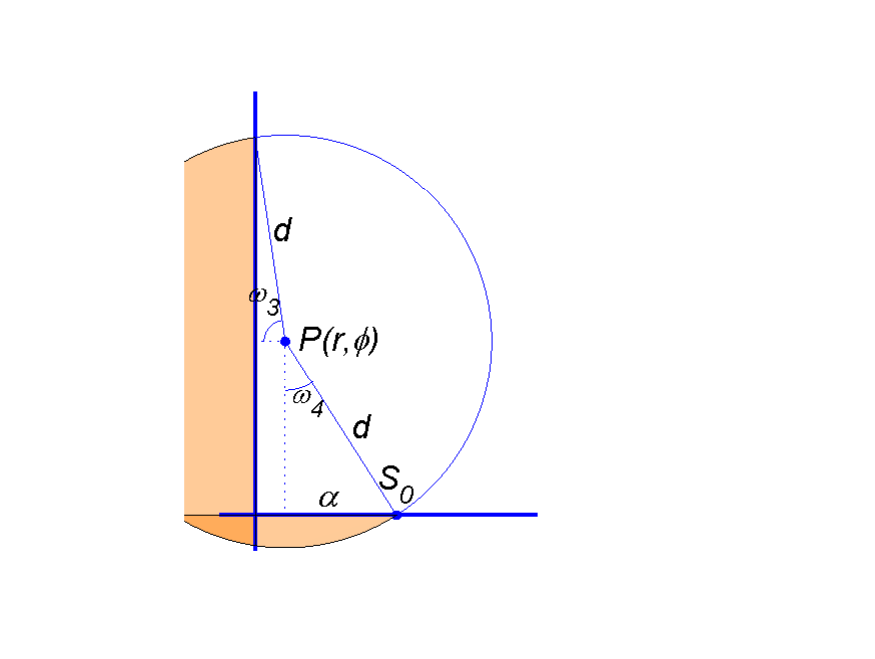}\label{fig:Integral3}} \hfil
\subfloat[]{\includegraphics[width=3.0in]{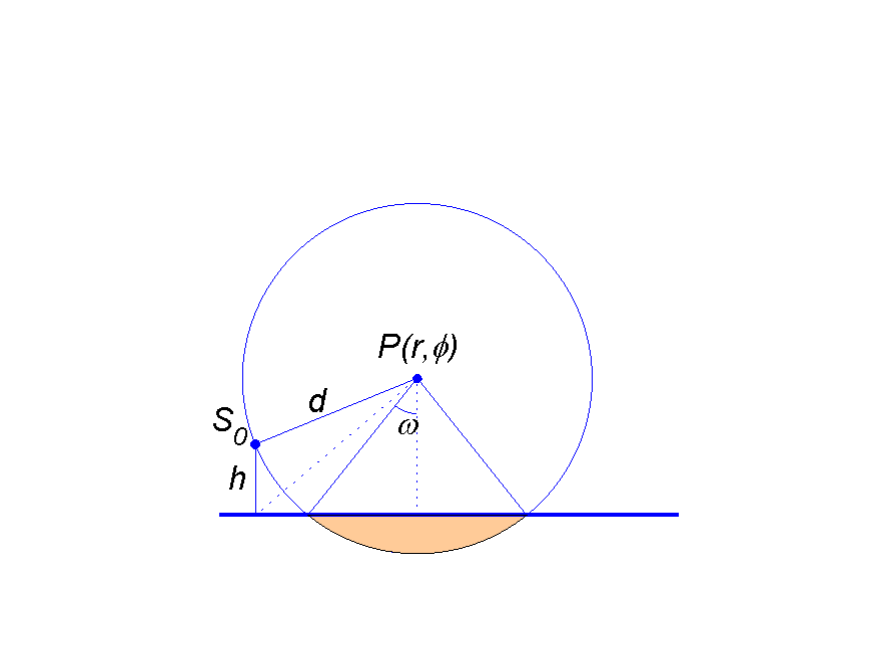}\label{fig:Integral4}}  
\caption{Coordinate systems and example illustrations of the void area of a point $P$ interior to Voronoi cell $\mathcal{C}_0$. In (a)$-$(c) the seed $S_0$ is located at distance $a$ from the corner of the quadrant. In (d), the seed $S_0$ is located at distance $h$ from the boundary of the half-plane. The void is the \emph{white-colored} part of the disk $D\!\left(P,d\!\left(P,S_0\right)\right)$; note the colored area clipped by the boundaries.}
\end{figure*}

In order to compute $\mathbb{E}\left\{A\right\}$, we should identify the probability that a point $P\in\mathbb{R}^2_+$ is interior to the cell $\mathcal{C}_0$, and integrate this probability over the quadrant. The point $P$ is interior to $\mathcal{C}_0$ when the seed $S_0$ lies on the circumference of the void of $P$. Since the underlying \ac{PPP} has unit intensity, this occurs with probability $e^{-V\left(P\right)}$, where  $V(P)\!=\!\left|D\left(P,d\left(P,S_0\right)\right)\cap\mathbb{R}^2_+\right|$. The mean cell area can be read as  
\[
\mathbb{E}\left\{A\right\}\!=\!\mathbb{E}\left[\int_{\mathbb{R}^2_+} \mathbbm{1}_{P\in\mathcal{C}_0}{\rm d}P \right]\!=\!\int_{\mathbb{R}^2_+}\mathbb{P}\left(P\in\mathcal{C}_0\right){\rm d}P \!=\! \int_{\mathbb{R}^2_+} e^{-V\left(P\right)}{\rm d}P, 
\]
where $\mathbbm{1}$ is the indicator function, equal to one for all points $P\in\mathcal{C}_0$ and zero otherwise. 

Let us assume that $S_0$ is located at the boundary of the quadrant and at distance $a$ from the corner. Given $a$, we separate between the following cases in the calculation of $V(P)$: 
\begin{itemize}
\item $r\geq a/2, \phi\leq\phi_1$, see Fig.~\ref{fig:Integral1}. For $\phi=\phi_1$, the void of $P$ becomes tangential to the boundary along the y-axis. In order to calculate $\phi_1$ we note that $d\!=\!r \cos \phi_1$. In addition, $d\!=\!\sqrt{r^2+a^2-2ar\cos\phi_1}$. Solving for the positive $\phi_1$, we end up with $\phi_1=\arccos\left(\frac{-a + \sqrt{2a^2+r^2}}{r}\right)$. For $\phi\leq\phi_1$, the boundary along the x-axis cuts some part of the disk $D\!\left(P,d\right)$. The angle $\omega$ in Fig.~\ref{fig:Integral1} can be calculated as  $\omega=\arccos\left(\frac{r \sin \phi}{d} \right)$, and the area of the void, denoted by $V_1$, 
is \begin{equation}
\label{eq:A1}
V_1 = \pi d^2 - \omega d^2 + r \sin\phi \left|r \cos\phi - a\right|.
\end{equation}

\item $r\geq a/2, \phi_1\leq\phi\leq\phi_2$, see Fig.~\ref{fig:Integral2}. For $\phi\!=\!\phi_2$, the edge of the disk $D\left(P,d\right)$ passes through the corner of the quadrant, and $d\!=\!r$. In addition, $d=\sqrt{r^2+a^2-2ar\cos\phi_2}$, hence,  $\phi_2=\arccos\left(\frac{a}{2r}\right)$. For $\phi_1\leq\phi\leq\phi_2$, both boundaries along the x- and y-axis determine the void, see Fig.~\ref{fig:Integral2}. In  Fig.~\ref{fig:Integral2}, $\omega_1=\omega$,  $\omega_2=\arccos\left(\frac{r \cos\phi}{d} \right)$, and the area of the void, $V_2$, is 
\begin{equation}
\label{eq:A2}
V_2 = \pi d^2 - \left(\omega_1+\omega_2\right) d^2 + r \sin\phi \left|r \cos\phi - a\right| + r d \cos\phi \sin\omega_2.
\end{equation}

\item $r\geq a/2, \phi_2\leq\phi\leq\pi/2$. In that case, see  Fig.~\ref{fig:Integral3}, the area of the void, denoted by $V_3$, can be calculated as the sum of a trapezium, a triangle and a circular domain with radius $d$ and angle $\left(\frac{3\pi}{2} -\omega_3-\omega_4\right)$, where $\omega_3=\omega_2$ and  $\omega_4=\omega$. Hence, 
\begin{equation}
\label{eq:A3}
V_3 = \frac{1}{2} r \sin\phi\left(r\cos\phi+a\right) + \frac{1}{2} r d \cos\phi \sin\omega_3 + \frac{ \frac{3\pi}{2} - \omega_3 - \omega_4}{2 \pi} \pi d^2.
\end{equation}

\item $r\leq a/2, \phi\leq \pi/2$. In that case, $\phi_1=\phi_2=0$, and the void  of $P$ always contains the corner of the quadrant in its interior. The area of the void is still given by equation~\eqref{eq:A3}. 
\end{itemize}
Finally, one has to sum up the four terms to consider all points in the quadrant. 
\begin{equation}
\label{eq:Mean}
\mathbb{E}\left\{A\right\} \!=\! \int\nolimits_0^{\phi_1}\!\!\!\!\! \int\nolimits_{\frac{a}{2}}^\infty\!\!\!\!\!\! e^{-V_1} r {\rm d}r {\rm d}\phi +\! \int\nolimits_{\phi_1}^{\phi_2}\!\!\!\int\nolimits_{\frac{a}{2}}^\infty\!\!\!\!\!\!e^{-V_2} r {\rm d}r {\rm d}\phi +\!\! \int\nolimits_{\phi_2}^{\frac{\pi}{2}}\!\!\! \int\nolimits_{\frac{a}{2}}^\infty\!\!\! e^{-V_3} r {\rm d}r {\rm d}\phi +\! \int\nolimits_{0}^{\frac{\pi}{2}}\!\!\!\int\nolimits_{0}^{\frac{a}{2}}\!\!\! e^{-V_3} r {\rm d}r {\rm d}\phi. 
\end{equation}
\begin{lemma} For a \ac{PVT} induced by a unit-intensity \ac{PPP} in the quadrant $\mathbb{R}^2_+$, the mean area of the cell $\mathcal{C}_0$ falling inside the quadrant is $\frac{\arccos\left(\frac{2}{\pi}\right)}{\sqrt{\pi^2-4}}$ when the seed $S_0$ is located at the corner. 
\begin{proof}
When the seed $S_0$ is located at the corner of the quadrant, one may substitute $a\!=\!0$, $\phi_1\!=\!0$ and $\phi_2\!=\!\pi/2$ in equation~\eqref{eq:Mean}. Therefore the area of the void of $P$ is essentially computed from equation~\eqref{eq:A2} after substituting $d=r,\, \omega_1=\frac{\pi}{2}-\phi$ and $\omega_2=\phi$. Finally we get $V\!\left(P\right)\!=\!\left(\frac{\pi}{2}\!+\!\sin\!\left(2\phi\right)\right)r^2$, and the mean cell area can be read as 
\begin{equation}
\label{eq:MeanCorner}
\mathbb{E}\left\{A\right\} = \displaystyle \int\nolimits_0^{\pi/2}\!\! \int\nolimits_{0}^\infty \!\! e^{-r^2\left( \frac{\pi}{2} + \sin\left( 2\phi\right) \right)} r {\rm d}r {\rm d}\phi = \frac{\arccos\left(\frac{2}{\pi}\right)}{\sqrt{\pi^2-4}}\approx 0.36351.
\end{equation}
\end{proof}
\end{lemma}
\begin{figure*}
\centering
\subfloat[\label{fig:Bound1}]{\includegraphics[width=0.325\textwidth]{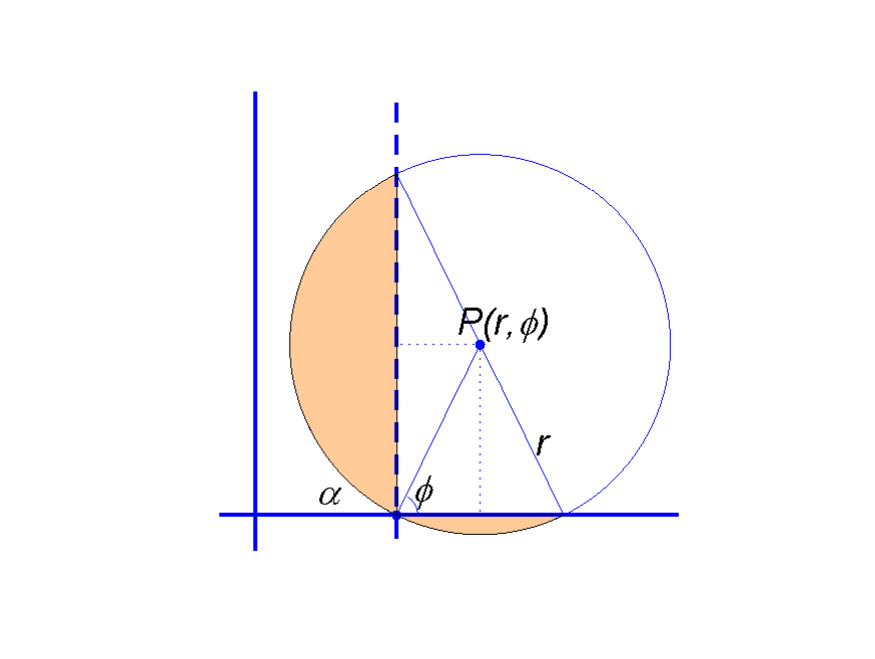}} \hfil  
\subfloat[\label{fig:Bound2}]{\includegraphics[width=0.325\textwidth]{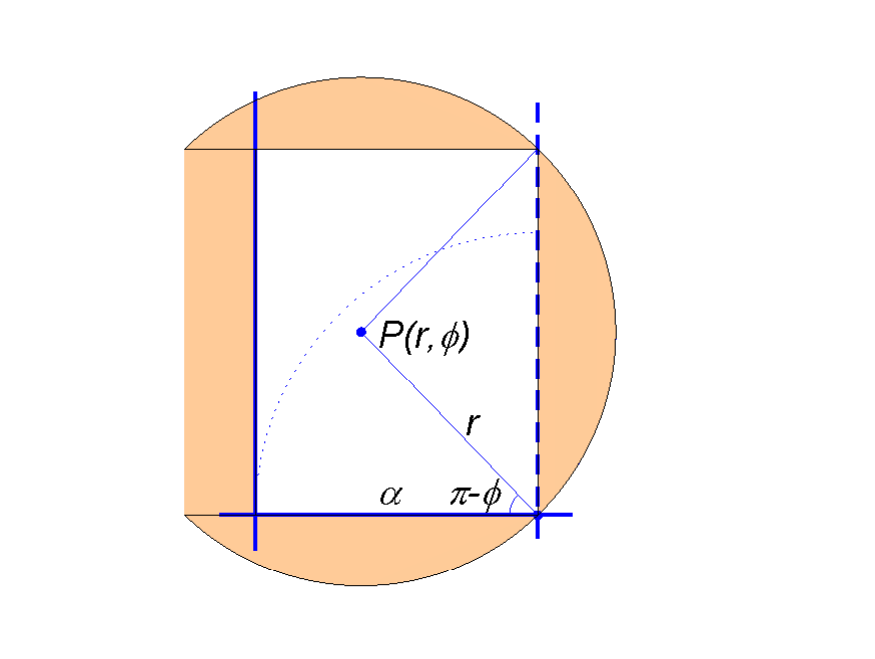}} \hfil 
\subfloat[\label{fig:Bound3}]{\includegraphics[width=0.325\textwidth]{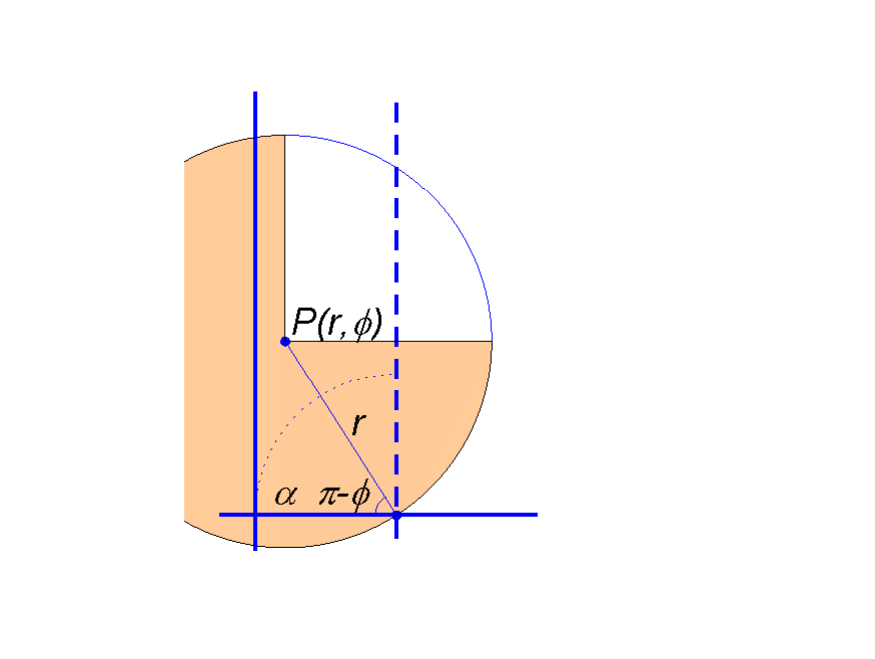}} 
\caption{Example illustrations in the calculation of the upper bound for the mean area $\mathbb{E}\!\left\{A\right\}$ in Lemma~\ref{lem:2}.}
\label{fig:Bound}
\end{figure*}
\begin{lemma} For a \ac{PVT} induced by a unit-intensity \ac{PPP} in the quadrant $\mathbb{R}^2_+$, the mean area of the cell $\mathcal{C}_0$ falling inside the quadrant is less than unity when the seed $S_0$ is located at the boundary. 
\label{lem:2}
\begin{proof}
In order to compute an upper bound (which is less than unity) of equation~\eqref{eq:Mean} for arbitrary $a\!\geq\!0$, we change the coordinate system so that the seed $S_0$ becomes the origin, and we construct lower bounds for the areas of the void of $P$, which can be evaluated in closed- and/or semi-closed form. We will consider all points $P\in\mathbb{R}^2_+$. Note that the coordinates of the boundaries of the quadrant is $x\!=\!-a$ and $y\!=\! 0$ in the new coordinate system, see Fig.~\ref{fig:Bound}.

When $r\!\geq\! 0, \, 0\!\leq\!\phi\!\leq\!\frac{\pi}{2}$, see Fig.~\ref{fig:Bound1}, we construct a lower bound for the area of the void considering that the y-axis, $x\!=\! 0$, see the dashed line in Fig.~\ref{fig:Bound1}, is a boundary. Thus, the mean cell area $\mathbb{E}\!\left\{A\right\}$ due to these points $P\!\left(r,\phi\right)$ is actually upper-bounded by equation~\eqref{eq:MeanCorner}. 

In order to bound the area of the void for $0\!\leq\! r \!\leq\! a, \, \frac{\pi}{2}\!\leq\!\phi\!\leq\! \pi$, see Fig.~\ref{fig:Bound2}, we use the area of the rectangle with sides $a$ and $2r\sin\phi$. Thus, the mean cell area $\mathbb{E}\!\left\{A\right\}$ due to these points $P\!\left(r,\phi\right)$ is upper-bounded by 
\[ 
\displaystyle \int\nolimits_0^a\!\!\int\nolimits_{\frac{\pi}{2}}^\pi \!e^{-2ar\sin \phi} r {\rm d}r {\rm d}\phi = -\frac{\pi}{4}\mathbf{M}_1\!\left(2a^2\right),
\]
where $\mathbf{M}_\nu\!\left(x\right)$ is the modified Struve function of the second kind, $\mathbf{M}_\nu\!\left(x\right)\!=\! \mathbf{L}_\nu\!\left(x\right)-I_\nu\!\left(x\right)$, where $\mathbf{L}_\nu\!\left(x\right)$ is the modified Struve function of the first kind, see~\cite[pp.~498]{Abramowitz1972}, and $I_\nu\!\left(x\right)$ is the modified Bessel function of the first kind, see~\cite[pp.~374]{Abramowitz1972}.

For the remaining points, i.e, $r\!>\! a, \, \frac{\pi}{2}\!\leq\!\phi\!\leq\!\pi- \arccos\left(\frac{a}{r}\right)$, see Fig.~\ref{fig:Bound3}, a lower bound on the area of the void is obtained by considering just the quarter of the disk $D\!\left(P,r\right)$. Hence, the mean cell area $\mathbb{E}\!\left\{A\right\}$ due to these points $P\!\left(r,\phi\right)$ is upper-bounded by  
\[
\displaystyle \int\nolimits_a^\infty\!\!\int\nolimits_{\frac{\pi}{2}}^{\pi-\arccos\left(\frac{a}{r}\right)} \!e^{-\frac{\pi}{4}r^2} r {\rm d}\phi {\rm d}r = e^{-\frac{a^2 \pi}{4}} - {\text{Erfc}}\!\left(\frac{a\sqrt{\pi}}{2}\right),
\] 
where ${\text{Erfc}}\!\left(x\right)\!=\!\frac{2}{\sqrt{\pi}}\int_x^\infty{e^{-t^2}{\rm d}t}$ is the complementary error function.

\noindent 
After summing up the contributions from the three parts of the quadrant we get 
\begin{equation}
\label{eq:Bounda}
\mathbb{E}\!\left\{A\right\} \!<\! e^{-\frac{a^2 \pi}{4}} -  {\text{Erfc}}\!\left(\frac{a\sqrt{\pi}}{2}\right) -\frac{\pi}{4} \mathbf{M}_1\!\left(2a^2\right) + \frac{\arccos\left(\frac{2}{\pi}\right)}{\sqrt{\pi^2-4}}. 
\end{equation}

The upper bound in~\eqref{eq:Bounda} can be evaluated at arbitrary precision,  and it is less than unity for all $a\!\geq \! 0$, see the red line in Fig.~\ref{fig:MeanA}. As $a\!\rightarrow\!\infty$, the Struve function  converges to $\lim_{a\rightarrow\infty}\mathbf{M}_1\!\left(2a^2\right)\!=\! -\frac{2}{\pi}$, and the bound converges to $\frac{1}{2}\!+\!\frac{\arccos\left(\frac{2}{\pi}\right)}{\sqrt{\pi^2-4}}$.
\end{proof}
\end{lemma}

\begin{lemma} For a \ac{PVT} induced by a unit-intensity \ac{PPP} in the half-plane, the mean area of cell $\mathcal{C}_0$ within the half-plane is less than unity when the seed $S_0$ is located at the boundary.
\begin{proof} 
We take a coordinate system where the seed $S_0$ is the origin. The area of the void for points with coordinates $r\geq 0$,  $0\leq\phi\leq \pi/2$ can be calculated using equation~\eqref{eq:A1} after substituting $d\!=\!r$, $a\!=\!0$ and $\omega\!=\!\frac{\pi}{2}-\phi$. After some straightforward calculation we get 
\begin{equation}
\label{eq:MeanEdge}
\mathbb{E}\!\left\{A\right\} = \displaystyle 2\! \int\nolimits_0^{\pi/2} \!\!\!\int\nolimits_{0}^\infty \!\!\!\!\! e^{-r^2\left( \frac{\pi}{2} + \phi + \sin\left( \phi\right) \cos\left( \phi\right) \right)} r {\rm d}r {\rm d}\phi \!=\! \int\nolimits_0^{\pi/2}\!\!\!\!\!\frac{2 {\rm d}\phi}{\pi\!+\!2\phi\!+\!\sin\left(2\phi\right)} \approx 0.61082, 
\end{equation}
where we have multiplied by $2$ to account for the angles $\pi/2 \leq\phi\leq \pi$. 

\noindent 
One may also note that $\mathbb{E}\!\left\{A\right\} \!<\! \int\nolimits_0^{\frac{\pi}{2}}\!\frac{2 {\rm d}\phi}{\pi\!+\!2\phi} \!=\! \log\!\left(2\right)\!<\! 1$. Another way to prove that $\mathbb{E}\!\left\{A\right\} \!<\! 1$ is to take the limit of the bound in~\eqref{eq:Bounda} as $a\!\rightarrow\!\infty$, resulting to $\mathbb{E}\!\left\{A\right\} \!<\! \frac{1}{2}\!+\!\frac{\arccos\left(\frac{2}{\pi}\right)}{\sqrt{\pi^2-4}}\!<\! 1$.
\end{proof}
\end{lemma}
\begin{remark} A rather loose lower bound to equation~\eqref{eq:Mean} is obtained after neglecting the impact of boundaries on the area of the disks $D\!\left(P,d\right)$, and substituting $V_1\!=\!V_2\!=\!V_3\!=\!\pi d^2$ in equation~\eqref{eq:Mean}. Finally,  $\mathbb{E}\left\{A\right\}>\frac{1}{4}\left(1+{\text{Erf}\left(a\sqrt{\pi} \right)} \right) \forall a\!\geq\! 0$,  where ${\text{Erf}}\left(x\right)\!=\!\frac{2}{\sqrt{\pi}}\int_0^x{e^{-t^2}{\rm d}t}$ is the error function.
\label{rem:1}
\end{remark}

The computation of the mean cell area $\mathbb{E}\!\left\{A\right\}$ for varying $a$ using~\eqref{eq:Mean} is validated in Fig.~\ref{fig:MeanA}. One may also find there the lower bound, see Remark~\ref{rem:1}, and the upper bound,  see~\eqref{eq:Bounda}. For large $a$, the mean converges to the value given in~\eqref{eq:MeanEdge}. For small $a$, e.g., $a\!\leq\!\frac{1}{2}$, the vertical boundary reduces significantly the mean cell area. For intermediate values of $a$, e.g., $1\!\leq\! a\! \leq \!2$, the mean cell area is large when the cell $\mathcal{C}_0$ contains also the corner of the quadrant in its interior.
\begin{figure*}
\centering
\subfloat[\label{fig:MeanA}]{\includegraphics[width=0.5\textwidth]{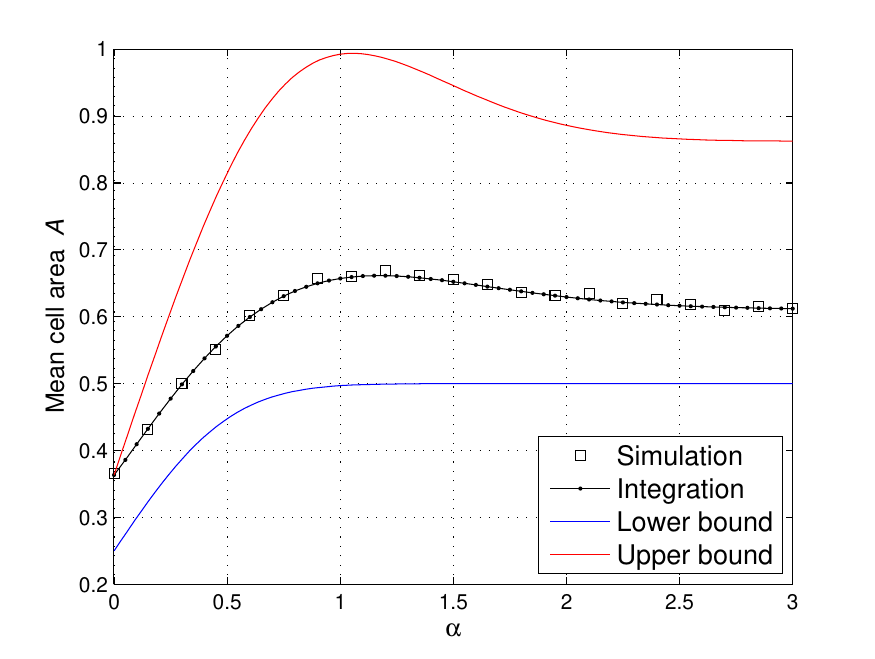}} \hfil 
\subfloat[\label{fig:MeanH}]{\includegraphics[width=0.5\textwidth]{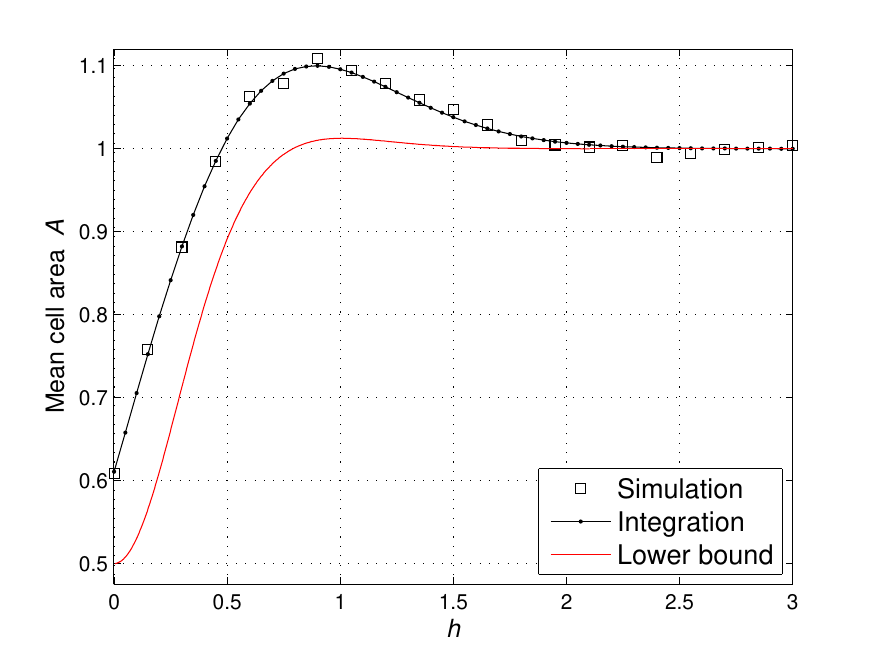}}
\caption{The integral-based calculation in equations~\eqref{eq:Mean} and~\eqref{eq:Mean2} is verified by simulations. $10\, 000$ simulation runs per marker. Mean cell area $\mathbb{E}\!\left\{A\right\}$ w.r.t. (a) the distance $a$ from the corner of the quadrant, (b) the distance $h$ from the boundary of the half-plane. The intensity of the \ac{PPP} is $\lambda\!=\! 1$. In the simulations, we consider a square with side $L\!=\! 10$. The \ac{PVT} and the area of the cell $\mathcal{C}_0$ falling inside the square are simulated using MatLab toolboxes. Given the set of seeds $S_i,i\!=\!0,1,2,\ldots$, the algorithm identifies the vertices of the tessellation, the boundary intersection points, and calculates the area of the polygon within the square associated with the cell $\mathcal{C}_0$. In the simulations for varying $h$, the coordinates of the seed $\mathcal{S}_0$ are $\left(\frac{L}{2},h\right)$.}
\label{fig:Mean}
\end{figure*}

Let us now assume that the seed $S_0$ is located at distance $h$ from the boundary of the half-plane, see Fig.~\ref{fig:Integral4}. In order to simplify the integration, the origin of the coordinate system is the point at the boundary nearest to $S_0$, thus the polar coordinates of $S_0$ become $\left(h,\frac{\pi}{2}\right)$. Following similar steps used to obtain  equations~\eqref{eq:A1}$-$\eqref{eq:Mean}, one can show that the mean cell area is 
\begin{equation}
\label{eq:Mean2}
\mathbb{E}\left\{A\right\} = 2\int\nolimits_0^{\phi_0} \!\!\!\int\nolimits_{h/2}^\infty\!\!\! e^{-V_1} r {\rm d}r {\rm d}\phi + 2\int\nolimits_{\phi_0}^{\pi/2} \!\!\!\int\nolimits_{h/2}^\infty\!\!\! e^{-V_2} r {\rm d}r {\rm d}\phi + 2\int\nolimits_{0}^{\pi/2}\!\!\! \int\nolimits_{0}^{h/2}\!\!\! e^{-V_1} r {\rm d}r {\rm d}\phi, 
\end{equation}
where $V_1\!=\!\left(\!\pi\!-\!\omega\!+\!\frac{\sin\left(2\omega\right)}{2} \!\right)\!d^2$, $V_2\!=\!\pi d^2$, $d\!=\!\sqrt{r^2\!+\!h^2\!-\!2hr\sin\!\phi}$, $\phi_0\!=\!\arcsin\!\left(\!\frac{-h \!+\! \sqrt{2h^2+r^2}}{r}\!\right)$, $\omega\!=\!\arccos\left(\!\frac{r \sin \phi}{d} \!\right)$, and the factor $2$ has been added to account for angles $\pi/2 \leq\phi\leq \pi$. 
\begin{figure*}
\centering
\subfloat[\label{fig:Bound1h}]{\includegraphics[width=.3\textwidth]{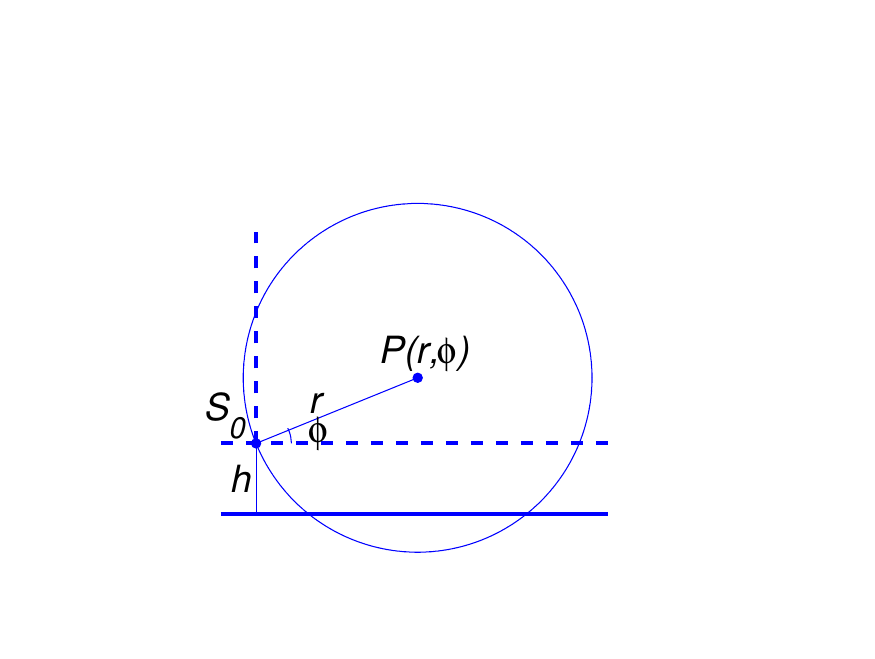}} \hfil 
\subfloat[\label{fig:Bound2h}]{\includegraphics[width=.3\textwidth]{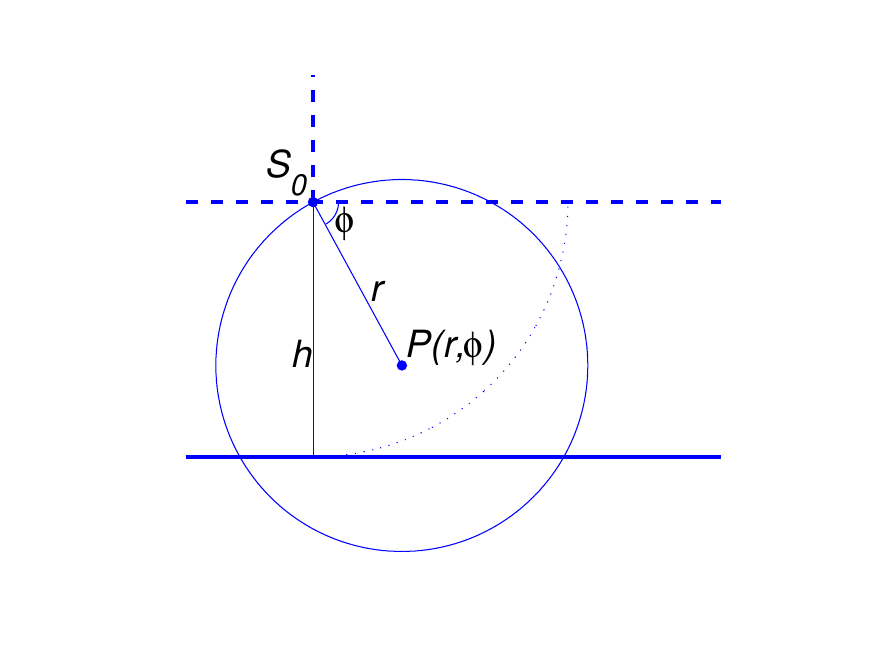}} \hfil 
\subfloat[\label{fig:Bound3h}]{\includegraphics[width=.3\textwidth]{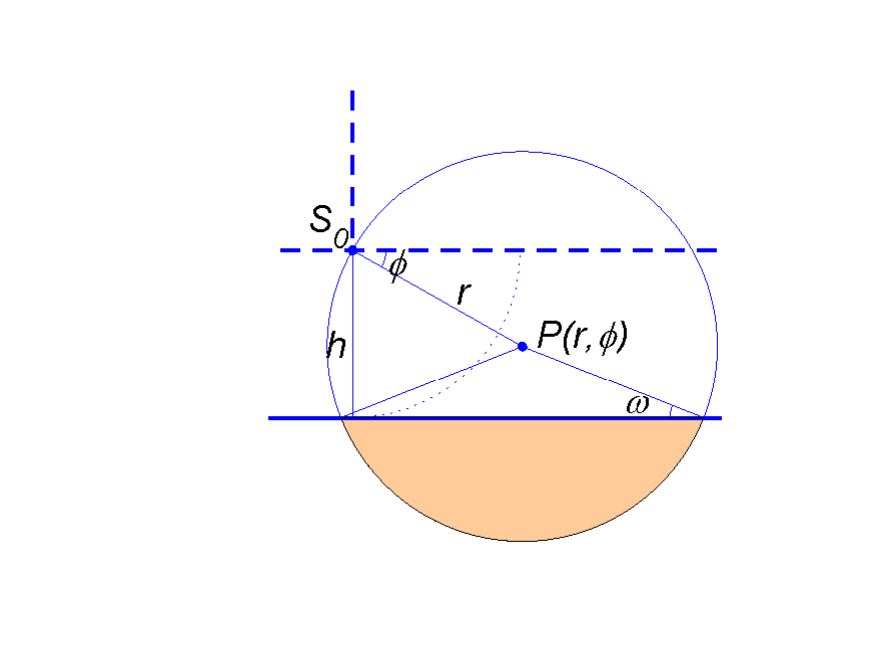}} 
\caption{Example illustrations in the calculation of the lower bound for the mean cell size in Lemma~\ref{lem:3}.}
\end{figure*}
\begin{lemma} For a \ac{PVT} induced by a unit-intensity \ac{PPP} in the half-plane, the mean area of the cell $\mathcal{C}_0$ falling in the half-plane can be larger than unity when the seed $S_0$ lies close to boundary.
\label{lem:3}
\begin{proof}
We look for a lower bound on the mean area $A$ which is larger than unity. First, we note that the lower bound obtained by setting $V_1\!=\!V_2\!=\!\pi d^2$ in equation~\eqref{eq:Mean2} is equal to $\frac{1}{2} \left(1+{\text{Erf}\left(h\sqrt{\pi} \right)} \right) \forall h\!\geq\! 0$. This is increasing in $h$ becoming unity as $h\!\rightarrow\!\infty$, thus cannot be used to claim mean cell area larger than unity. In order to obtain another lower bound to equation~\eqref{eq:Mean2}, we start by changing the coordinate system so that the seed $S_0$ becomes the origin. In the new system, see the dashed lines in Fig.~\ref{fig:Bound1h}$-$\ref{fig:Bound3h}, the coordinates of the boundary is $y\!=\!-h$. Then, we will construct appropriate upper bounds for the area of the void. Note that the void of a general point $P$ in the quadrant is now the intersection of the half-plane and of the disk centered at $P$ with radius equal to the distance between $P$ and the nearest seed(s). 

When $r\!\geq\! 0, \, 0\!\leq\! \phi\!\leq\! \frac{\pi}{2}$, see Fig.~\ref{fig:Bound1h}, we may neglect the impact of boundary on the area of the void with negligible approximation error, thus the mean cell area $\mathbb{E}\!\left\{A\right\}$ due to these points $P\!\left(r,\phi\right)$ is lower-bounded by  
\[ 
2 \displaystyle \int\nolimits_0^\infty\!\!\int\nolimits_0^{\frac{\pi}{2}} \!e^{-\pi r^2} r {\rm d}r {\rm d}\phi = \frac{1}{2}.
\]

When $r\!\leq\! h, \, -\frac{\pi}{2} \!\leq\! \phi\!\leq\! 0$, see Fig.~\ref{fig:Bound2h}, we still neglect the impact of boundary on the area  of the void. This approximation may introduce non-negligble error for the points with radii $\frac{h}{2}\!\leq\! r \!\leq\! h$.
\[ 
2 \displaystyle \int\nolimits_0^{h}\!\!\int\nolimits_{-\frac{\pi}{2}}^0 \!e^{-\pi r^2} r {\rm d}r {\rm d}\phi = \frac{1}{2}\left( 1-e^{-\pi h^2}\right).
\]

Finally, for the remaining points $r \!\geq\! h, \, -\arcsin\!\left(\frac{h}{r}\right)\!\leq\! \phi\!\leq\! 0$, see Fig.~\ref{fig:Bound3h}, the area of the void is  $V\!\left(P\right) \!=\! \left(\frac{\pi}{2}+\omega+\cos\omega\sin\omega\right)r^2$, where $\omega\!\left(\phi\right)\!=\! \arcsin\!\left(\frac{h}{r}+\sin\phi\right)$. Due to the fact that $\frac{1}{2}\sin\!\left(2x\right)\!<\!x$ for  $x\!\geq\! 0$, the area can be upper-bounded by $V\!\left(P\right)\!\leq\! \left(\frac{\pi}{2}+2\omega\right)r^2$. For $r\!\geq\! h, -\frac{\pi}{2}\!\leq\!-\arcsin\!\left(\frac{h}{r}\right)\!\leq  \phi\!\leq\! 0$, the function $\omega\!\left(\phi\right)$ is increasing in $\phi$ with positive second derivative. Hence, $V\!\left(P\right)\!\leq\!\left(\frac{\pi}{2} +2\left(\arcsin\left(\frac{h}{r}\right) + \phi \right) \right)r^2$.  Therefore the contribution of these points to the mean cell area can be lower-bounded as 
\[ 
\int_h^\infty\!\!\!\int_{-\arcsin\left(\frac{h}{r}\right)}^0\!\! e^{-\left(\frac{\pi}{2} +2\left(\arcsin\left(\frac{h}{r}\right) + \phi \right) \right)r^2} \! r{\rm d}\phi {\rm d}r = \frac{1}{2}{\text{Ei}}\!\left(\frac{\pi h^2}{2}\right) - \int_h^\infty \!\frac{1}{r}e^{-\left(\frac{3\pi}{2}-2\arccos\left(\frac{h}{r}\right)\right)r^2} \! {\rm d}r, 
\] 
where ${\text{Ei}}\!\left(x\right)\!=\!\int_x^\infty \frac{e^{-t}}{t}{\rm d}t, x\!>\!0$ is the exponential integral.

In order to lower bound the right-hand side of the equation above, we need to upper bound the second term. A rather trivial upper bound is obtained using a piecewise function to upper-bound  $\arccos\!\left(\frac{h}{r}\right)$, i.e, $\frac{\pi}{3}$ for $h\!\leq\! r\!\leq\! 2h$ and $\frac{\pi}{2}$ for $r\!>\! 2h$.  
\[ 
\int_h^\infty \!\frac{1}{r}e^{-\left(\frac{3\pi}{2}-2\arccos\left(\frac{h}{r}\right)\right)r^2} \! {\rm d}r \!<\! \frac{1}{2}\left({\text{Ei}}\!\left(\frac{5\pi h^2}{6}\right) - {\text{Ei}}\!\left(\frac{10\pi h^2}{3}\right) + {\text{Ei}}\!\left(2\pi h^2\right) \right)
\]

\noindent 
After summing up the contributions from the three parts of the half-plane we get 
\begin{equation}
\label{eq:Boundh}
\begin{array}{ccl}
\mathbb{E}\left\{A\right\} &>& 1 - \frac{1}{2} e^{-\pi h^2} + \frac{1}{2}\Big( {\text{Ei}}\!\left(\frac{10\pi h^2}{3}\right) - {\text{Ei}}\!\left(\frac{5\pi h^2}{6}\right) + {\text{Ei}}\!\left(\frac{\pi h^2}{2}\right)-{\text{Ei}}\!\left(2\pi h^2\right)\Big).
\end{array}
\end{equation}

The right-hand side of~\eqref{eq:Boundh} can be evaluated at arbitrary precision. When the distance $h$ to the boundary is around $h\!=\! 1$, we observe mean cell sizes larger than unity, see Fig.~\ref{fig:MeanH}.  
\end{proof}
\end{lemma}
\begin{remark} 
In Fig.~\ref{fig:Mean}, we see that the integral-based calculation matches quite well the simulation results even for a moderate average number of points, $\lambda L^2\!=\! 100$, inside the square where the simulated \acp{PVT} are generated. Note that the probability that the cell $\mathcal{C}_0$ touching opposite sides of the square is at most $\exp\!\left(-\lambda \pi L^2/8\right)$, thus negligible for our parameter settings. 
\end{remark}

\section{Second moment of cell area $\mathbb{E}\!\left\{A^2\right\}$}
\label{sec:Variance}
In order to calculate the second moment of the cell area, one has to consider two points $P_1\!\left(r_1,\phi_1\right), P_2\!\left(r_2,\phi_2 \right)$ interior to the cell $\mathcal{C}_0$ and average over their locations.  
\[
\mathbb{E}\left\{A^2\right\}\!=\!\int_{\mathbb{R}^2_+\times \mathbb{R}^2_+ }\mathbb{P}\left(P_1,P_2\in\mathcal{C}_0\right){\rm d}P_1{\rm d}P_2 \!=\! \int_{\mathbb{R}^2_+\times \mathbb{R}^2_+} e^{-V\left(P_1,P_2\right)}{\rm d}P_1{\rm d}P_2, 
\]
where $V\!\left(P_1,P_2\right)\!=\! \left|\left(D\left(P_1,d\left(P_1,S_0\right)\right) \cup D\left(P_2,d\left(P_2,S_0\right)\right)\right) \cap\mathbb{R}^2_+\right|$ is the area of the intersection of two disks and the quadrant, and the points $S_0,P_1,P_2$ cannot be collinear. 

In the infinite plane, the calculation of the second moment using integral-based methods can be found in~\cite{Brakke1986,Hayen2002}. The computation of $V\!\left(P_1,P_2\right)$ in a bounded domain is cumbersome. Nevertheless, when the seed $S_0$ is fixed either at the corner of the quadrant, $a\!=\!0$, or at the boundary of the half-plane, $h\!=\!0$, the second moment can be calculated using few integral terms. 
\begin{lemma}
For a \ac{PVT} induced by a unit-intensity \ac{PPP} in the quadrant $\mathbb{R}^2_+$, the second moment of the area of $\mathcal{C}_0$ falling inside the quadrant when the seed $S_0$ is located at the corner is 
\[
\mathbb{E}\left\{A^2\right\} \!=\! \int\nolimits_0^{\frac{\pi}{2}}\!\int\nolimits_{\theta-\frac{\pi}{2}}^\theta\!\int\nolimits_{-\omega_1}^{\frac{\pi}{2}-\theta}\!\!\frac{f\left(\omega_1,\omega_2\right) {\rm d}\omega_2 {\rm d}\omega_1 {\rm d}\theta}{V_1^2} + \! 2 \int\nolimits_{-\frac{\pi}{2}}^{0}\!\int\nolimits_{-\frac{\pi}{2}}^{\theta}\!\int\nolimits_{-\omega_1}^{\frac{\pi}{2}}\!\!\frac{f\left(\omega_1,\omega_2\right) {\rm d}\omega_2 {\rm d}\omega_1 {\rm d}\theta}{V_2^2},  
\]
where $V_1 = \frac{2\theta + \sin\left(2\left(\theta-\omega_1\right)\right) + \sin\left(2 \omega_1\right)}{2\cos^2\omega_1}  +  \frac{\pi - 2\theta + \sin\left(2\left(\theta+\omega_2\right)\right) + \sin\left(2 \omega_2\right)}{2\cos^2\omega_2}$, $V_2\!=\!\frac{\pi + 2\sin\left(2\left(\theta+\omega_2\right)\right)}{2\cos^2\omega_2}$, and $f\left(\omega_1,\omega_2\right) \!=\! \frac{\sin\left( \omega_1+\omega_2\right)}{\cos^3\omega_1 \cos^3\omega_2 }$.
\label{lem:4}
\begin{proof}
\begin{figure*}
\centering
\subfloat[\label{fig:Integral21}]{\includegraphics[width=0.4\textwidth]{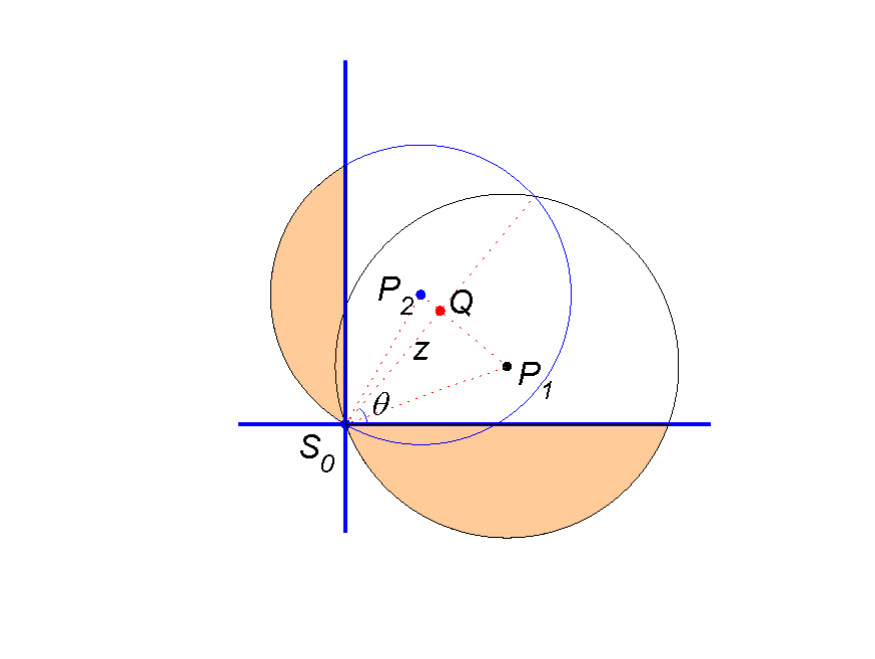}} \hfil  
\subfloat[\label{fig:Integral22}]{\includegraphics[width=0.4\textwidth]{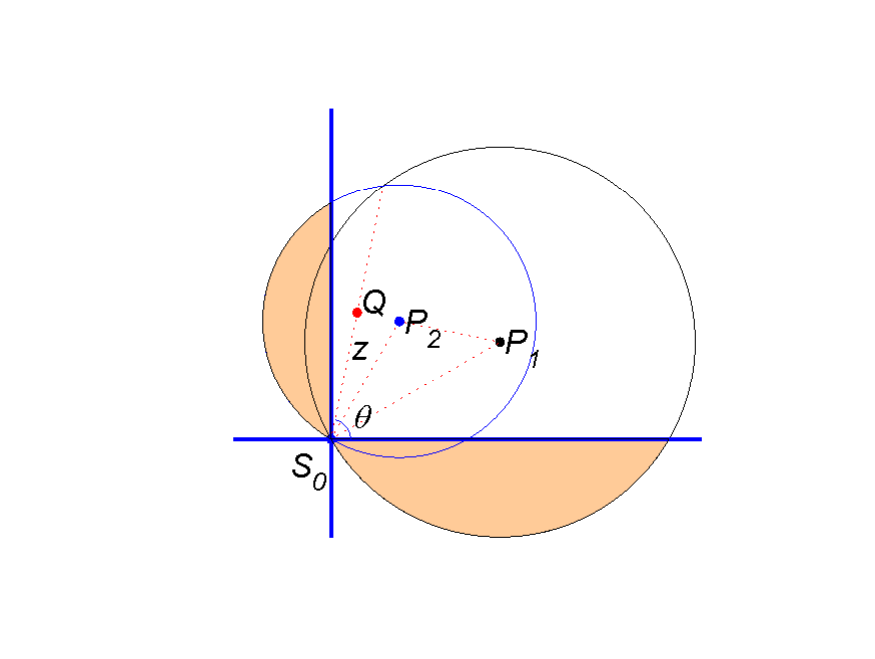}} \hfil
\subfloat[\label{fig:Integral23}]{\includegraphics[width=0.4\textwidth]{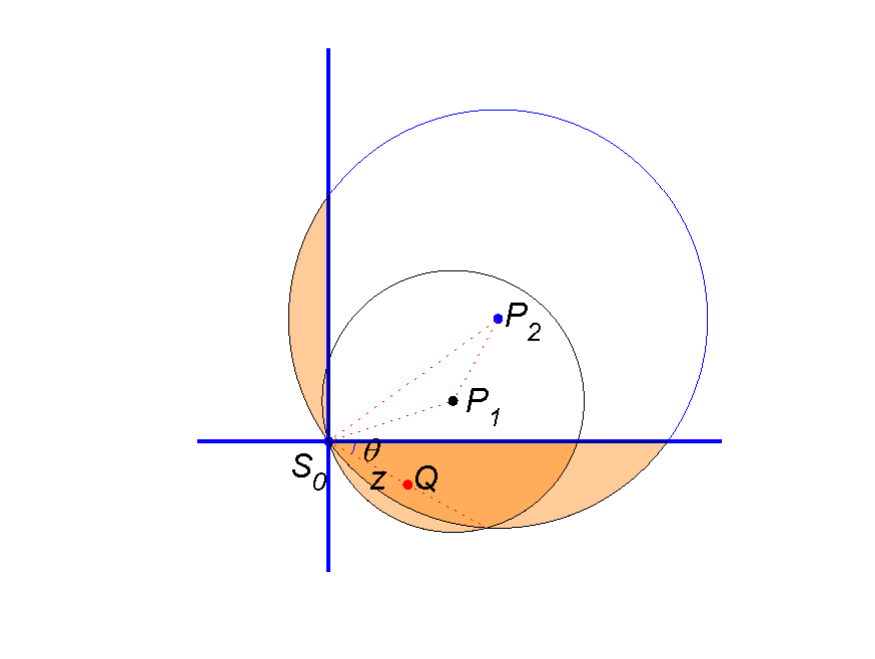}} \hfil 
\subfloat[\label{fig:Integral24}]{\includegraphics[width=0.4\textwidth]{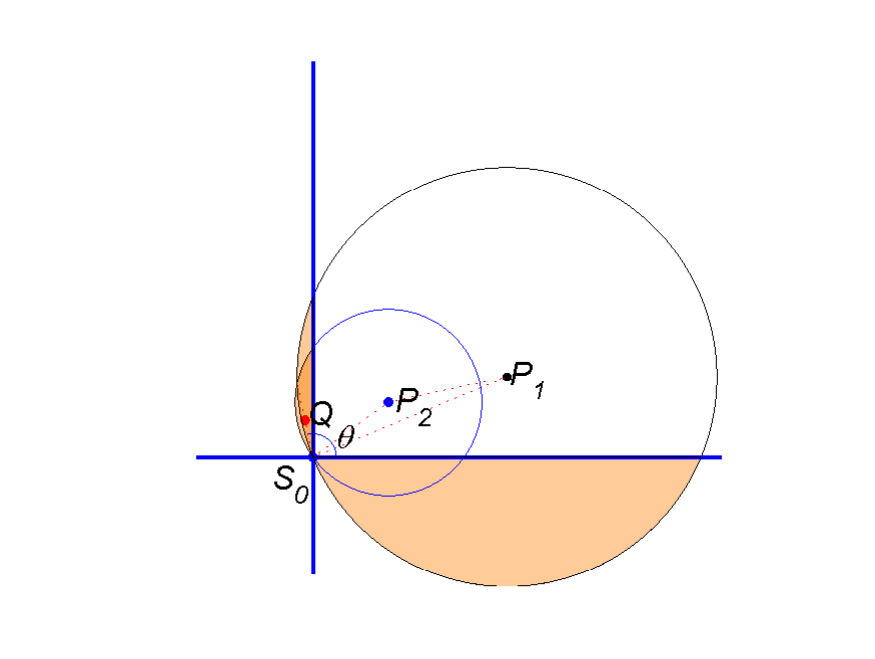}} 
\caption{Coordinate systems and example illustrations of the void of two points $P_1, P_2$ when the seed $S_0$ is located at the corner.}
\label{fig:IntegralStd}
\end{figure*}

We transform the coordinate system as follows: $\left(z,\theta\right)$ are the polar coordinates of the point $Q$ which is the intersection point of the line passing through $P_1,P_2$ and its perpendicular line passing through the origin $S_0$, $\omega_1$ is the angle $QS_0P_1$ measured clockwise, and $\omega_2$ is the angle $QS_0P_2$ measured counter-clockwise, see Fig.~\ref{fig:IntegralStd} for example illustrations. The transformation can be read as $\phi_1 \!=\!\theta \!-\! \omega_1$, $\phi_2\!=\!\theta \!+\! \omega_2$, $r_1\!=\!\frac{z}{\cos\omega_1}$ and $r_2\!=\!\frac{z}{\cos\omega_2}$. The determinant of the Jacobian matrix of the transformation is $|J| \!=\! z^3 f\left(\omega_1,\omega_2\right)$, where $f\left(\omega_1,\omega_2\right) \!=\! \frac{\sin\left( \omega_1+\omega_2\right)}{\cos^3\omega_1 \cos^3\omega_2 }$. 

Due to the fact that $r_1\!=\!\frac{z}{\cos\omega_1}$ and $r_2\!=\!\frac{z}{\cos\omega_2}$, the area of the void $V\!\left(P_1,P_2\right)$ can be written as $V_j z^2$, where $V_j$ is the area of the void normalized for $z\!=\!1$. After integrating the probability that the seed $S_0$ is at the edge of the void, $e^{-V_j z^2}$, we have $\mathbb{E}\!\left\{A^2\right\}\!=\!\int e^{-V_j z^2} z^3 f\left(\omega_1,\omega_2\right) {\rm d} P_1 {\rm d} P_2$. Integrating over $z\!\geq\!0$, we get  $\mathbb{E}\!\left\{A^2\right\}\!=\!\int\frac{f\left(\omega_1,\omega_2\right)}{2 V_j^2} {\rm d}\omega_2 {\rm d}\omega_1 {\rm d}\theta$. 

\noindent 
In the infinite plane, i.e., in the bulk of the domain, the normalized area of the void is 
\[
V=\frac{\pi + 2\omega_1 + \sin\left(2\omega_1\right)}{2\cos^2\omega_1} \!+\! \frac{\pi + 2\omega_2 + \sin\left(2\omega_2\right)}{2\cos^2\omega_2}, 
\]
resulting to $\mathbb{E}\!\left\{A^2\right\}\!\approx\!1.28$~\cite{Brakke1986, Hayen2002}. 

When the seed $S_0$ is located at the corner, even though both points $P_1,P_2$  are located within the quadrant, the angle $\theta$ can take values in $\left[-\frac{\pi}{2},\pi\right]$. The range of the variables $\omega_1,\omega_2$ depend on the quadrant where the point $Q$ lies. Therefore the computation of the void can be divided into three parts. When the point $Q$ lies in the upper-right quadrant, the angles $\omega_1, \omega_2$ could be positive or negative. Example illustrations are in Fig.~\ref{fig:Integral21}, where the angle $\omega_2$ is positive and in Fig.~\ref{fig:Integral22}, where $\omega_2$ is negative. In both figures, $\omega_1$ is positive. In order to calculate the area of the void, we take the disk generated by $P_1$ and subtract: (i) the shaded area under the x-axis, and (ii) the part of the disk at the left of the line passing through $S_0$ and $Q$. In a similar manner, we can calculate the contribution to the void due to the point $P_2$. After summing up we get 
\[
V_1 = \frac{2\theta + \sin\left(2\left(\theta-\omega_1\right)\right) + \sin\left(2 \omega_1\right)}{2\cos^2\omega_1}  +  \frac{\pi - 2\theta + \sin\left(2\left(\theta+\omega_2\right)\right) + \sin\left(2 \omega_2\right)}{2\cos^2\omega_2}. 
\]

When $\theta$ is negative, e.g., in Fig.~\ref{fig:Integral23}, $\omega_1$ becomes always negative and $\omega_2$ always positive. In Fig.~\ref{fig:Integral23}, we see that the point $P_1$ can be ignored, and the area of the void, denoted by $V_2$, can be calculated based on $P_2$, i.e., $V_2\!=\!\frac{\pi + 2\sin\left(2\left(\theta+\omega_2\right)\right)}{2\cos^2\omega_2}$. Finally, when $\frac{\pi}{2}\leq \theta \leq \pi$, see Fig.~\ref{fig:Integral24}, the area of the void depends only on $P_1$, and $V_3\!=\!\frac{\pi + 2\sin\left(2\left(\theta-\omega_1\right)\right)}{2\cos^2\omega_1}$. Due to symmetry, negative angles, $\theta\leq 0$ and angles larger than $\frac{\pi}{2}$ give equal contributions. In addition, every integral term must be multipled by two to consider each pair of points twice and the Lemma is proved. 
\end{proof}
\end{lemma} 
\begin{lemma} For a \ac{PVT} induced by a unit-intensity \ac{PPP} in the half-plane, the second moment of the size of the cell $\mathcal{C}_0$ when the seed $S_0$ is located at the boundary is
\[
\begin{array}{ccc}
\mathbb{E}\left\{A^2\right\} \!&\!=\!& \displaystyle\!  \int\nolimits_0^{\frac{\pi}{2}}\!\!\int\nolimits_{\theta-\frac{\pi}{2}}^\theta\!\int\nolimits_{-\omega_1}^{\frac{\pi}{2}-\theta}\frac{2f\!\left(\omega_1,\omega_2\right) \!{\rm d}\omega_2 {\rm d}\omega_1 {\rm d}\theta}{V_1^2} + \int\nolimits_0^{\frac{\pi}{2}}\!\!\int\nolimits_{\theta-\frac{\pi}{2}}^\theta\!\int\nolimits_{\frac{\pi}{2}-\theta}^{\frac{\pi}{2}}\!\!\frac{2f\!\left(\omega_1,\omega_2\right) \!{\rm d}\omega_2 {\rm d}\omega_1 {\rm d}\theta}{V_2^2} + \\ 
& & \displaystyle \int\nolimits_0^{\frac{\pi}{2}}\!\!\int\nolimits_{-\frac{\pi}{2}}^{\theta-\frac{\pi}{2}}\!\!\int\nolimits_{-\omega_1}^{\frac{\pi}{2}}\!\!\frac{2f\left(\omega_1,\omega_2\right) {\rm d}\omega_2 {\rm d}\omega_1 {\rm d}\theta}{V_3^2} + \int\nolimits_{-\frac{\pi}{2}}^{0}\!\!\int\nolimits_{-\frac{\pi}{2}}^{\theta}\!\int\nolimits_{-\omega_1}^{\frac{\pi}{2}}\!\!\frac{2f\left(\omega_1,\omega_2\right) {\rm d}\omega_2 {\rm d}\omega_1 {\rm d}\theta}{V_4^2},
\end{array} 
\]
where 
\[
\begin{array}{ccl}
V_1 &\!=\!& \frac{2\theta + \sin\left(2\left(\theta-\omega_1\right)\right) + \sin\left(2 \omega_1\right)}{2\cos^2\omega_1}  \!+\! \frac{\pi+2\omega_2+\sin\left(2\omega_2\right) }{2\cos^2\omega_2} \\ 
V_2 &\!=\!& \frac{2\theta + \sin\left(2\left(\theta-\omega_1\right)\right) + \sin\left(2 \omega_1\right)}{2\cos^2\omega_1}  \!+\! \frac{2\pi-2\theta + \sin\left(2 \omega_2\right) - \sin\left(2 \left( \theta+\omega_2\right)\right)}{2\cos^2\omega_2} \\ 
V_3 &\!=\!& \frac{\pi + 2\omega_1 +  \sin\left(2 \omega_1\right)}{2\cos^2\omega_1}  \!+\! \frac{2\pi-2\theta + \sin\left(2 \omega_2\right) - \sin\left(2 \left( \theta+\omega_2\right)\right)}{2\cos^2\omega_2} \\ 
V_4 &\!=\!& \frac{\pi+2\theta+2\omega_2 +\sin\left(2\left(\theta+\omega_2\right)\right)} {2\cos^2\omega_2}.
\end{array}
\]
\label{lem:5}
\begin{proof}
\begin{figure*}
\centering
\subfloat[\label{fig:Integral25}]{\includegraphics[width=0.4\textwidth]{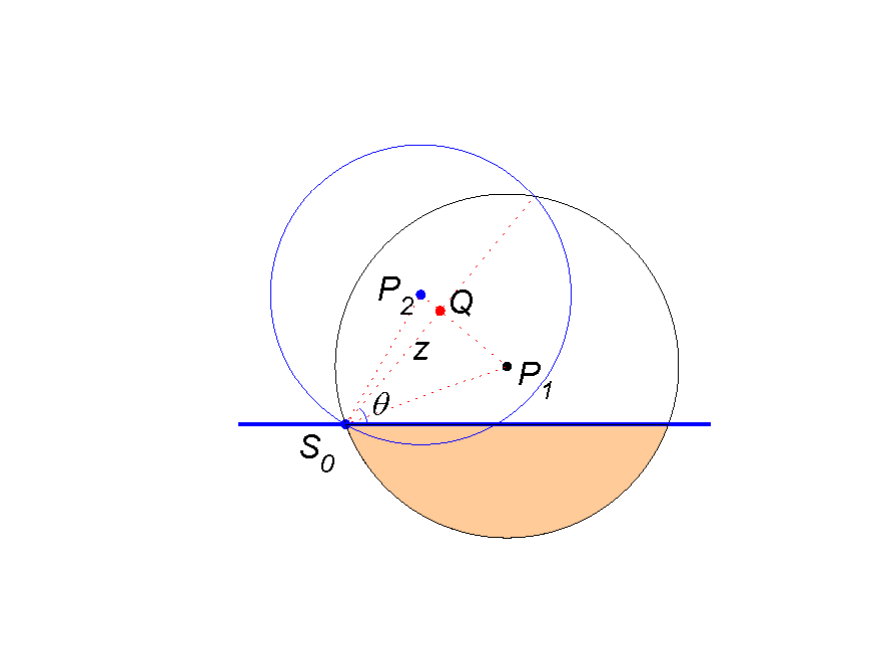}} \hfil 
\subfloat[\label{fig:Integral26}]{\includegraphics[width=0.4\textwidth]{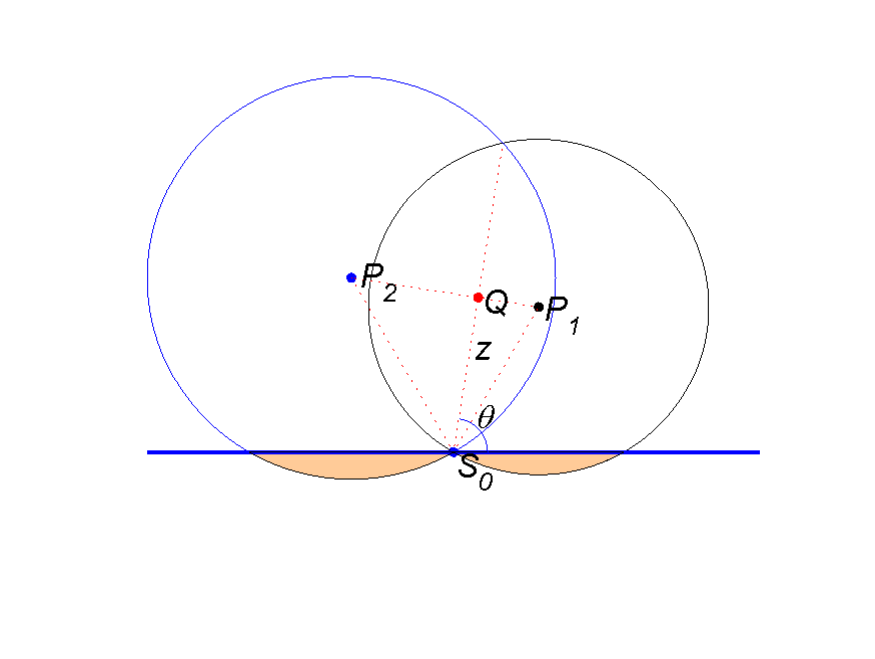}} \hfil 
\subfloat[\label{fig:Integral27}]{\includegraphics[width=0.4\textwidth]{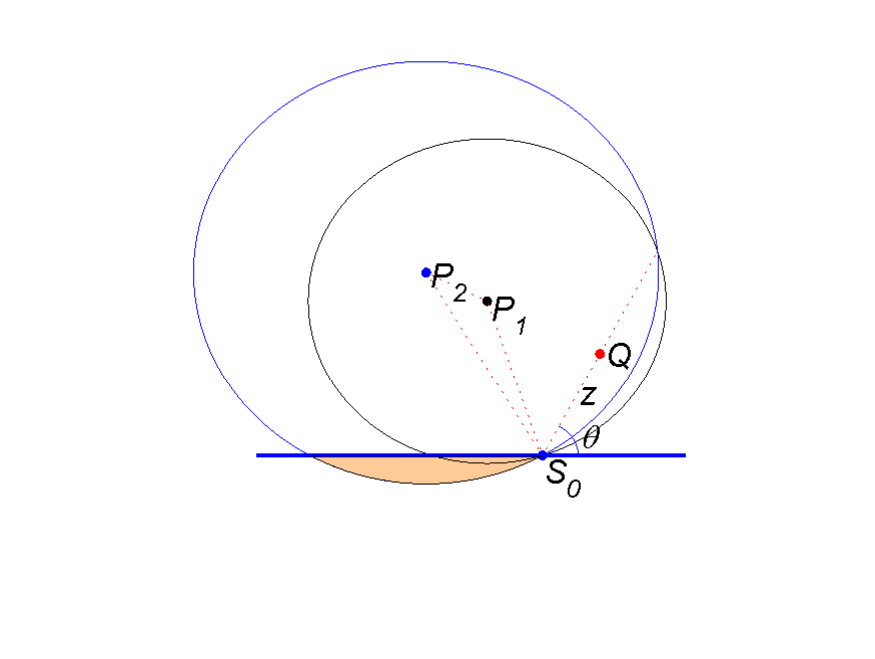}} \hfil
\subfloat[\label{fig:Integral28}]{\includegraphics[width=0.4\textwidth]{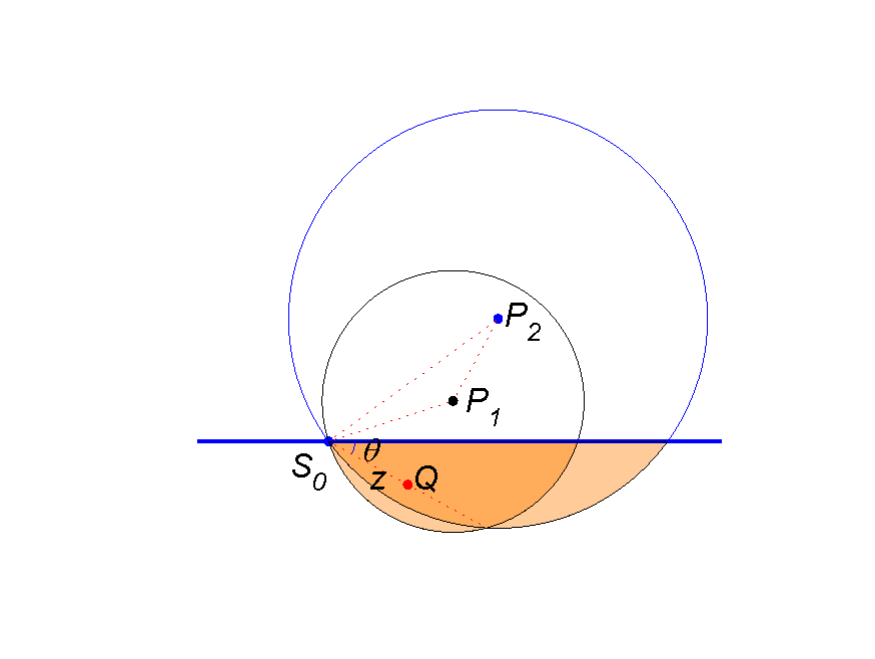}}
\caption{Coordinate systems and example illustrations of the void area around two interior points $P_1, P_2$ when the seed $S_0$ is located at the boundary of the half-plane.}
\label{fig:IntegralStd2}
\end{figure*}
We consider same coordinate system with Lemma~\ref{lem:4}, with the seed $S_0$ being the origin. We separate between angles $-\frac{\pi}{2}\leq\theta\leq\frac{\pi}{2}$ and $\frac{\pi}{2}\leq\theta\leq\frac{3\pi}{2}$. Due to symmetry, it is sufficient to carry out the computation only for $-\frac{\pi}{2}\leq\theta\leq\frac{\pi}{2}$. For $0\leq\theta\leq\frac{\pi}{2}$, all configurations of points $P_1,P_2$ can be divided into three cases: Both points are located at the upper-right quadrant,  $\left\{\phi_1\leq\frac{\pi}{2},\phi_2\leq\frac{\pi}{2}\right\}$, point $P_1$ is located at the upper-right and point $P_2$ at the upper-left quadrant, $\left\{\phi_1\leq\frac{\pi}{2},\frac{\pi}{2}\leq\phi_2\leq\pi\right\}$, and both points are located at the upper-left quadrant, $\left\{\frac{\pi}{2}\leq\phi_1\leq\pi,\frac{\pi}{2}\leq\phi_2\leq\pi\right\}$. An example  configuration for the first case is depicted in Fig.~\ref{fig:Integral25}, where the disk of $P_2$ is not anymore limited from a boundary along the y-axis, thus 
\[
V_1 \!=\! \frac{2\theta + \sin\left(2\left(\theta-\omega_1\right)\right) + \sin\left(2 \omega_1\right)}{2\cos^2\omega_1}  \!+\! \frac{\pi+2\omega_2+\sin\left(2\omega_2\right) }{2\cos^2\omega_2}.
\]

In the second case, both disks due to $P_1$ and $P_2$ are truncated from the boundary, see Fig.~\ref{fig:Integral26}. After some straightfoward calculation we get the size of the aggregate void, 
\[
V_2 \!=\! \frac{2\theta + \sin\left(2\left(\theta-\omega_1\right)\right) + \sin\left(2 \omega_1\right)}{2\cos^2\omega_1}  \!+\! \frac{2\pi-2\theta + \sin\left(2 \omega_2\right) - \sin\left(2 \left( \theta+\omega_2\right)\right)}{2\cos^2\omega_2}. 
\]
In the third case, see Fig.~\ref{fig:Integral27}, only the disk of $P_2$ is affected from the boundary
\[
V_3\!=\!\frac{\pi + 2\omega_1 +  \sin\left(2 \omega_1\right)}{2\cos^2\omega_1}  \!+\! \frac{2\pi-2\theta + \sin\left(2 \omega_2\right) - \sin\left(2 \left( \theta+\omega_2\right)\right)}{2\cos^2\omega_2}. 
\]

Finally, for $\theta\!<\!0$, see Fig.~\ref{fig:Integral28}, the void is determined only from $P_2$ and $V_4\!=\!\frac{\pi+2\theta+2\omega_2 +\sin\left(2\left(\theta+\omega_2\right)\right)} {2\cos^2\omega_2}$. After multiplying every term by four to consider angles $\frac{\pi}{2}\leq\theta\leq\frac{3\pi}{2}$ and to count every pair of points twice and summing up we get the result of the Lemma.
\end{proof}
\end{lemma}

After numerical integration we get $\mathbb{E}\left\{A^2\right\} \!=\! 0.23781$ in Lemma~\ref{lem:4} and  $\mathbb{E}\left\{A^2\right\} \!=\! 0.54508$ in Lemma~\ref{lem:5}. The associated mean values at the corner of the quadrant and at the boundary of the half-plane are given in equations~\eqref{eq:MeanCorner} and~\eqref{eq:MeanEdge} respectively. 

With the numerical calculation of the first two moments at hand, we may now select suitable distributions to approximate the \ac{PDF} of the cell area at different locations. The Gamma distribution has so far been widely used for the \ac{PDF} of the Voronoi cell area in the bulk, with two parameters in~\cite{Weaire1986,Pineda2004} and with three parameters in~\cite{Tanemura2003}. In~\cite{Weaire1986} the parameters $k,\nu$ of the Gamma \ac{PDF}, $\frac{x^{k-1}e^{-x/\nu} }{\nu^k\Gamma\left(k\right)}$, are selected equal to $k\!=\!\nu^{-1}\!=\!3.61$ using simulations, and in~\cite{Pineda2004}, they are selected equal to $k\!=\!\nu^{-1}\!=\!3.575$ using numerical integration. In~\cite{Tanemura2003}, the maximum likelihood function of the generalized Gamma \ac{PDF} is numerically maximized given $10^7$ data samples, resulting to the best up-to-date fit of Voronoi cell area \ac{PDF} in the bulk.
\begin{table}
\caption{Fitting the Gamma \ac{PDF} to the distribution of the area $A$ of the cell $\mathcal{C}_0$ using method of moments.}
\begin{center}
 \begin{tabular}{||c| c c c c||} 
\hline 
\label{table:Table1}
& $\mathbb{E}\!\left\{A\right\}$ & $\mathbb{V}\!{\text{ar}}\left\{A\right\}$ & $k$ & $\nu$ \\
\hline 
Corner & $0.36351$  & $0.10567$ & $1.25052$ & $0.29069$ \\ 
\hline
Edge & $0.61082$ & $0.17198$ & $2.16935$ & $0.28157$ \\ 
\hline
Bulk & $1$ & $0.28018$ & $3.56918$ & $0.28018$ \\ 
\hline
\end{tabular}
\end{center}
\end{table}
\begin{figure}
\centering
  \includegraphics[width=3.0in]{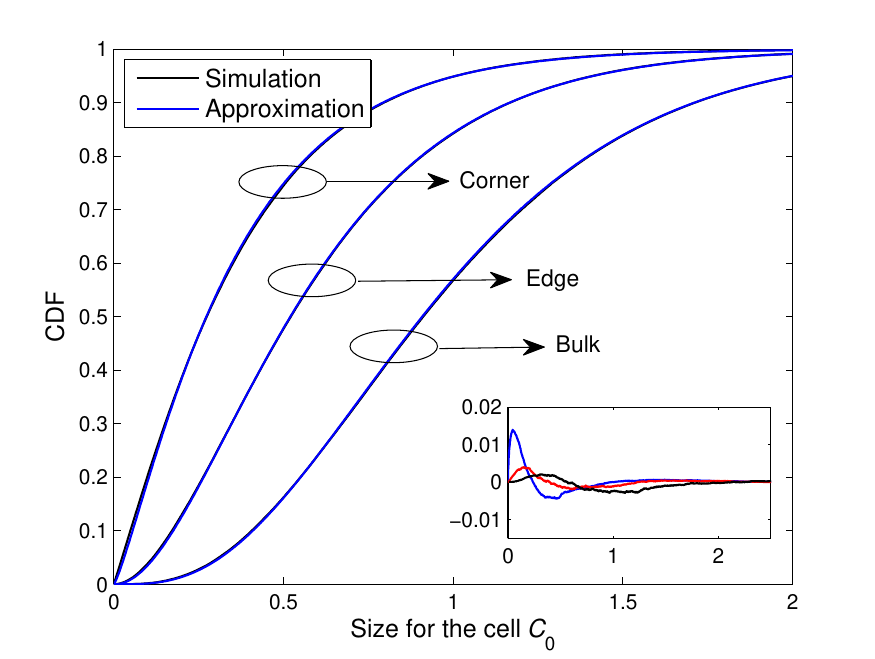}
 \caption{Distribution of the area of the Voronoi cell $\mathcal{C}_0$ falling inside the quadrant at different locations of the seed $S_0$ using simulations and the Gamma distribution with parameters available in Table~\ref{table:Table1}. In the inset, we depict the approximation error between the simulated and approximated \ac{CDF}, i.e., $F_{\text{sim}}\!-\! F_{\text{app}}$, where $F$ is the \ac{CDF}, at the corner (blue), at the edge (red) and in the bulk (black). It indicates that with the selected parameters the distributions are not Gamma.}
 \label{fig:Distributions}
\end{figure}
\begin{figure*}
\centering
\subfloat[\label{fig:ChiSquarePDFBulk}Bulk]{\includegraphics[width=.3\textwidth]{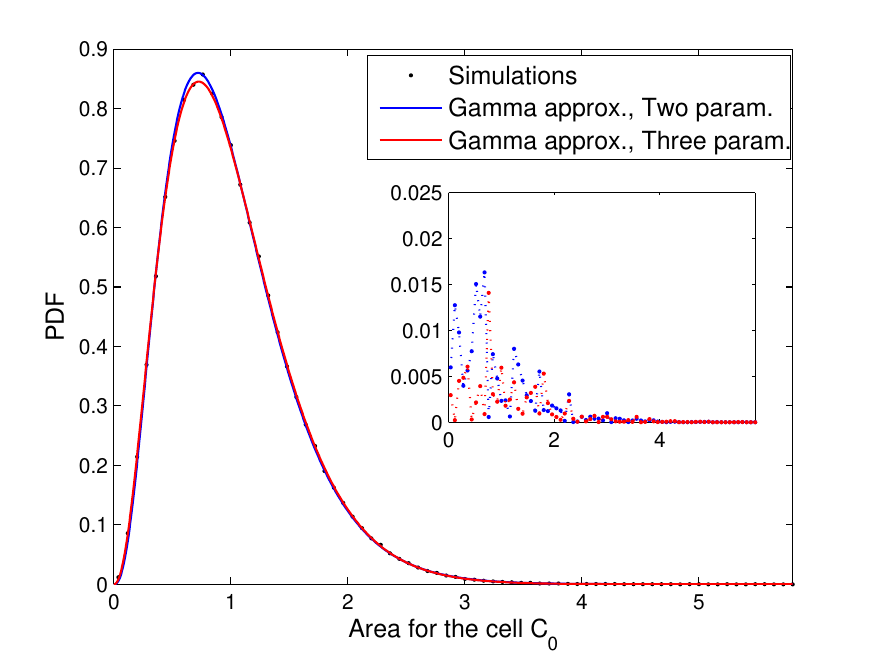}} \hfil  
\subfloat[\label{fig:ChiSquarePDFEdge}Edge]{\includegraphics[width=.3\textwidth]{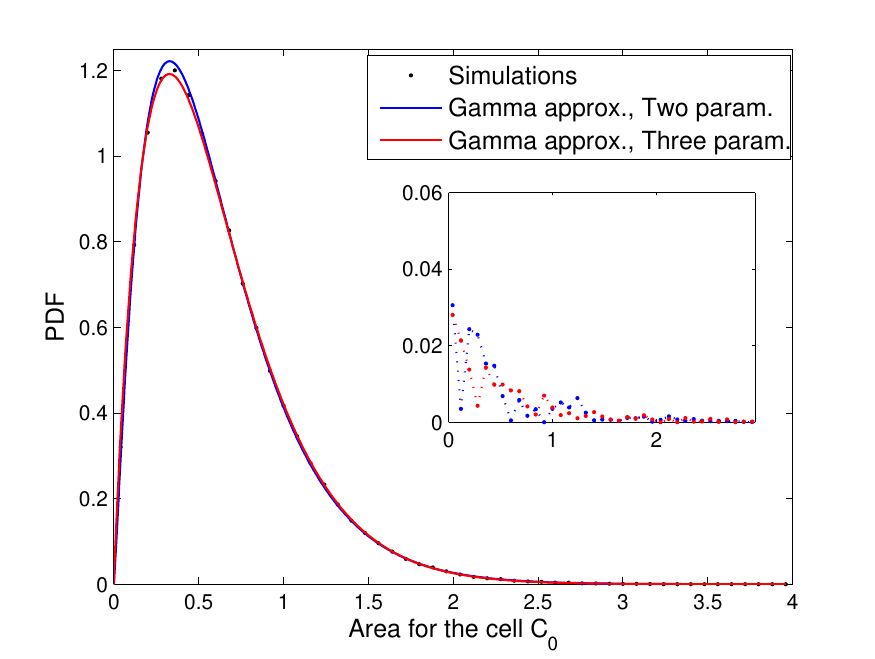}} \hfil
\subfloat[\label{fig:ChiSquarePDFCorner}Corner]{\includegraphics[width=.3\textwidth]{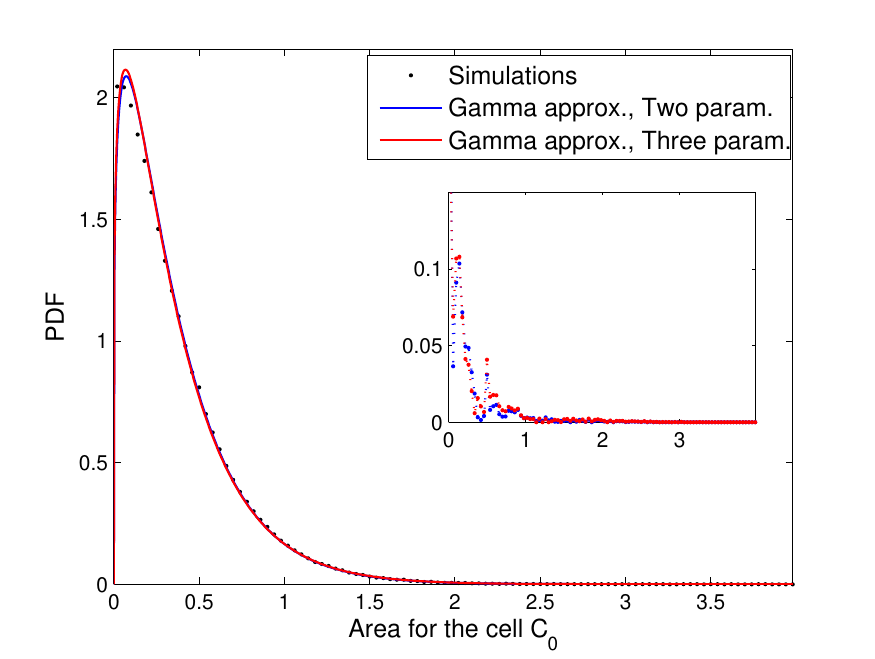}}
\caption{The simulated \acp{PDF} for the area of the Voronoi cell $\mathcal{C}_0$ and the Gamma approximation with parameters in Table~\ref{table:Table1}. $500\, 000$ \acp{PVT} simulations per location. In order to get smooth simulated points we consider class interval $0.08$ for the bulk and the edge, and $0.04$ for the corner. For the bulk the Chi-squared test statistic at $5 \%$ significance level is $441.9$ for the two-parameter Gamma and $109.3$ for the three-parameter Gamma, while the critical value with $55$ degrees of freedom is $73.3$. The estimates for the three-parameter Gamma \ac{PDF} are $a\!=\!1.09735, b\!=\!2.91894, c\!=\!3.25367$. For the edge, the Chi-squared statistic is approximately $239.9$ for the two-parameter Gamma and $192.7$ for the three-parameter Gamma, while the critical value with $45$ degrees of freedom is $61.6$. The estimates are $a\!=\!1.11495, b\!=\!2.95131, c\!=\!1.95004$. Finally, for the corner, the Chi-squared statistic is approximately $1076$ for the two-parameter Gamma and $964$ for the three-parameter Gamma, while the critical value with $75$ degrees of freedom is $96.2$. The estimates are $a\!=\!0.98351, b\!=\!3.44063, c\!=\!1.24211$. This is another justification that for the selected parameters the distributions are not Gamma. In the insets, the absolute difference $\left|f_{\text{sim}}\!-\! f_{\text{app}}\right|$ is depicted, where $f$ stands for the \ac{PDF}. The Chi-squared goodness-of-fit and the optimization using Quasi-Newton method have been done using MatLab toolboxes.}
\label{fig:ChiSquarePDF}
\end{figure*}

After fitting the moments to the two-parameter Gamma distribution, we end up with $k\!=\!\frac{\mathbb{E}\left\{A\right\}^2} {\mathbb{V}{\text{ar}}\left\{A\right\}}$ and $\nu\!=\!\frac{\mathbb{E}\left\{A\right\}}{k}$. The mean and the variance of the cell area $A$ as well as the parameters $k,\nu$ when the seed $S_0$ is located at the corner of the quadrant, at the edge of the half-plane, and in the infinite plane (i.e. in the bulk)  are summarized in Table~\ref{table:Table1}. We see that the parameter $k$ depends clearly on the location, while the parameter $\nu$ is not that sensitive. In the bulk, the values we get for $k,\nu$ are close to those of~\cite{Weaire1986,Pineda2004}. The parameterized Gamma distributions at the corner and at the edge are to the best of our knowledge new. In Fig.~\ref{fig:Distributions}, we have simulated $500\, 000$ \acp{PVT} over a square with side $L\!=\!10$ and \ac{PPP} intensity equal to unity. We see that the Gamma distribution with fitted mean and variance provides a good approximation for the \ac{CDF} of the area of $\mathcal{C}_0$ inside the square for all locations of the seed $S_0$. The simulated mean and variance at the corner are $0.36282$ and $0.10571$ respectively, while at the edge the related values are $0.61037$ and $0.17154$. The error between the simulations and the approximations, see the inset of Fig.~\ref{fig:Distributions}, and the Chi-squared goodness-of-fit test, see the caption in Fig.~\ref{fig:ChiSquarePDF}, indicate that the cell area distributions are not actually Gamma. In Fig.~\ref{fig:ChiSquarePDF}, we note that the main source of approximation error comes around the peak of the \ac{PDF} for all locations, especially at the corner. Nevertheless, the two-parameter Gamma approximations using method of moments should be adequate for use in many applications, see for instance the next section. Note that the three-parameter Gamma approximation depicted in Fig.~\ref{fig:ChiSquarePDF}, $\frac{a x^{c-1} e^{-b x^a} b^{c/a}}{\Gamma\left(c/a\right)}$, estimated by maximizing the log-likelihood function using Quasi-Newton method, see also~\cite{Tanemura2003}, provides better fit than the Gamma distribution. The improvement, as compared to the two-parameter Gamma, is significant only for the bulk. Also, the Chi-squared values at 5\% significance level indicate that the true distribution is not the three-parameter Gamma either.

\begin{figure*}
\centering
\subfloat[\label{fig:GridMean}]{\includegraphics[width=.5\textwidth]{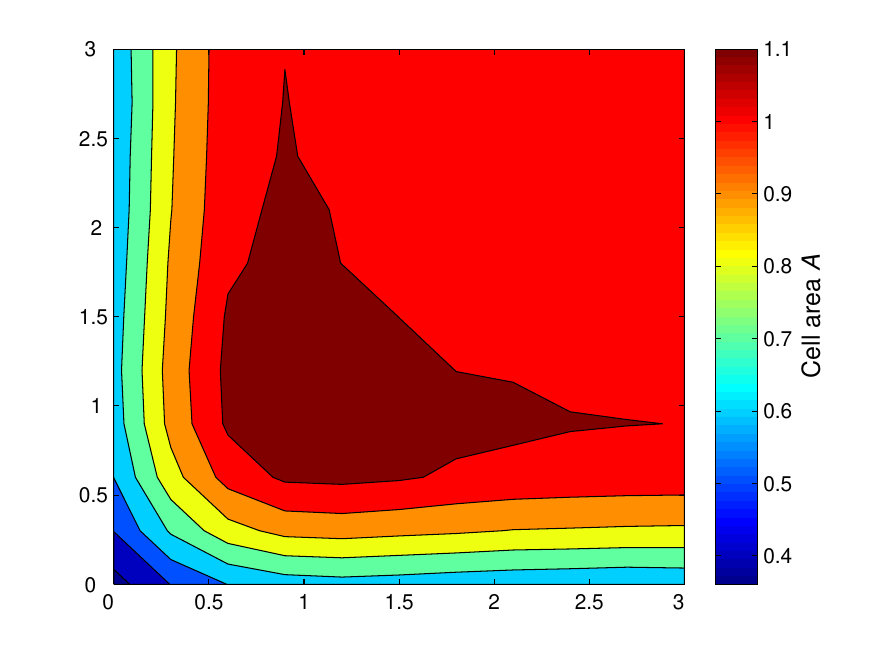}} \hfil  
\subfloat[\label{fig:GridStd}]{\includegraphics[width=.5\textwidth]{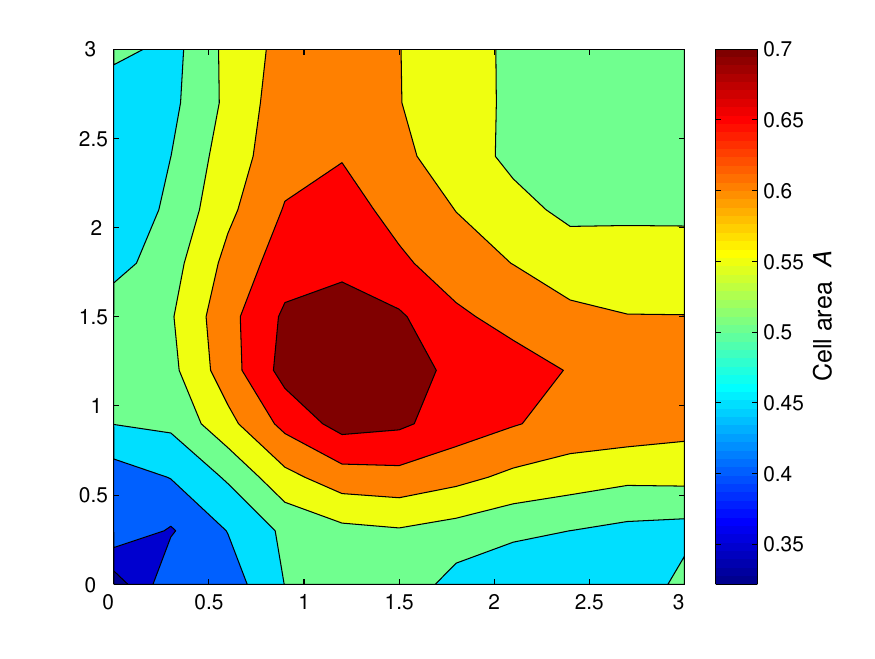}}
\caption{Contour plots for (a) the mean, and (b) the standard deviation of the cell area $A$ when the seed $S_0$ is located close to the corner of the quadrant. A grid of $11\times 11$ seeds is simulated with $10\, 000$ simulations per seed.}
\label{fig:Grid}
\end{figure*}

The numerical calculation of the mean and the variance of the cell area for seeds located close to the boundary and the corner of the quadrant involves very bulky integrals without getting any new insights. In Fig.~\ref{fig:Grid}, we have  simulated the contour plots for the mean and the standard deviation of cell area for a grid of seeds close to the corner of the quadrant. The coordinates of the seed $S_0$ are $\left(i\Delta x,j\Delta y\right)$, where $\Delta x\!=\!\Delta y\!=\! 0.3$ and $i=0,1,\ldots, 10, j=0,1,\ldots, 10$. As expected, the mean and the variance of the cell area are maximized when the seed is located close to the boundary and also close to the corner of the quadrant. In addition, we see in Fig.~\ref{fig:GridMean} that the mean cell area  converges quickly to unity as we move towards the bulk.  

\section{Application to physical layer security}
\label{sec:Secrecy}
In modern wireless communications systems, the confidentiality of information between a mobile user and the base station is maintained with authentication and cryptography. The main idea of cryptographic techniques is to share a secret key between the communication parties, which is used to encrypt source information at the transmitter and extract the information messages at the receiver. Even if an eavesdropper manages to decode the transmitted packets, it is very unlikely to extract any useful information from them, unless it possesses the decryption key, see~\cite[Chapter 1]{Wang2016} for more details. 

Physical layer security without exchanging secret keys was first proposed by Wyner~\cite{Wyner1975}. It refers to the protection of information messages against eavesdropping using the uniqueness of the wireless channel between the sender and the receiver~\cite{Trappe2015}. Physical layer security  would be well-suited for devices with light computational power, e.g., in certain types of wireless sensor networks (WSNs), where conventional cryptography-based techniques fail to adapt due to their high complexity that incurs a high power cost~\cite{Trappe2015}.

Let us consider an entity $B_1$ that wants to send a message to entity $B_2$. The message is protected against the $i$-th eavesdropper $E_i$, if the eavesdropper fails to extract useful information from the message it receives. We will assume that $B_1$ succeeds to send the message in a secure manner, if the distance between $B_1$ and $B_2$ is smaller than the distance between $B_1$ and the eavesdropper closest to $B_1$, i.e., $d\left(B_1,B_2\right)<d\left(B_1,E_i\right) \forall i$. This  distance-based criterion for secure connectivity is quite fundamental as it corresponds to the case without fading in the wireless channel, no interference, equal noise power levels at the legitimate users and the eavesdroppers, and secrecy rate threshold equal to zero~\cite{Pinto2012}. In~\cite{Koufos2019}, we have shown that boundaries can enhance physical layer security for a fixed wireless link in the presence of a single eavesdropper without the above assumptions. 

\begin{figure*}
\centering
\subfloat[\ac{AP} at the bulk]{\includegraphics[width=.5\textwidth]{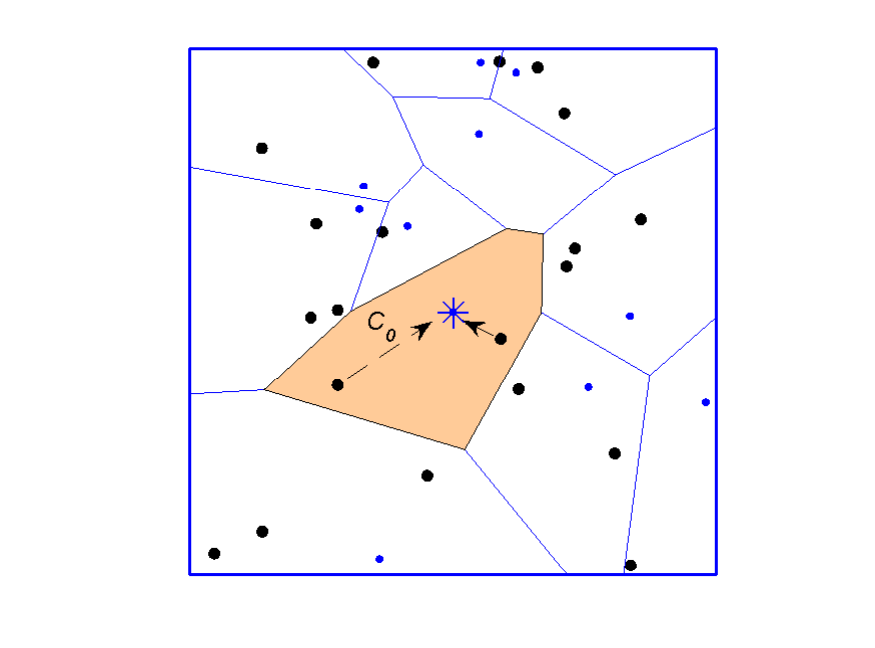}\label{fig:EavesBulk}} \hfil  
\subfloat[\ac{AP} at the boundary]{\includegraphics[width=.5\textwidth]{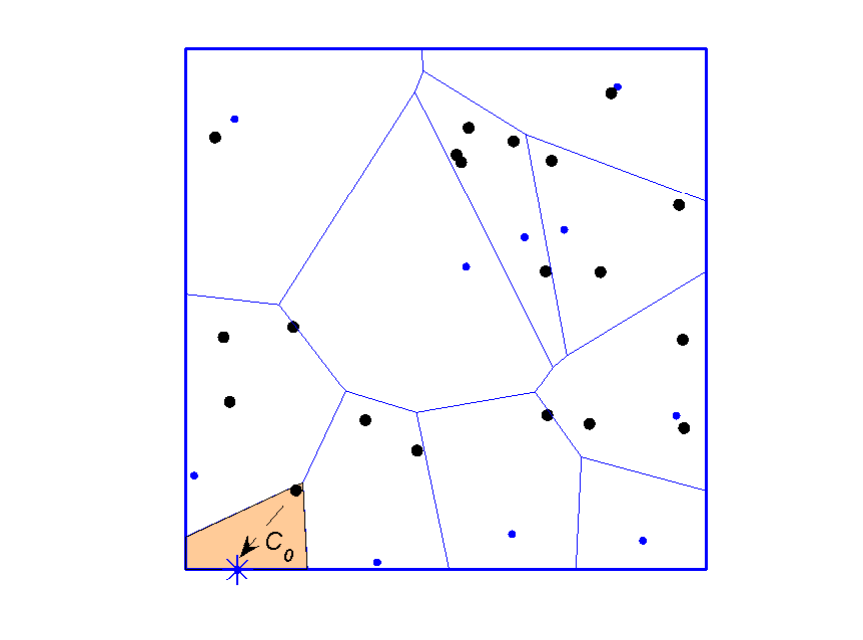}\label{fig:EavesBoundary}}
\caption{Example illustration for Voronoi tessellations generated by the \ac{PPP} of the eavesdroppers (blue dots) and the \ac{AP} (blue asterisk), $\Pi_e \cup \left\{S_0\right\}$. The number of legitimate users (black dots) with secure connection to the \ac{AP} located in the bulk is two. The corresponding number at the boundary is one. The part of the cell $\mathcal{C}_0$ falling inside the deployment domain is colored.}
\label{fig:Eaves}
\end{figure*}

We consider two independent and homogeneous \acp{PPP}; one for the legitimate users, $\Pi_l$, and another for the eavesdroppers, $\Pi_e$, with intensities $\lambda_l$ and $\lambda_e$ respectively. In addition, we place a node $S_0$ at a fixed location, either at the corner of the quadrant or at the edge of the half-plane. For instance, this could be the location of an \ac{AP} where all sensors (legitimate users) want to transmit measurement data. We would like to study the distribution of the number of legitimate users with a secure connection to and from the \ac{AP}, over the ensemble of all possible realizations of $\Pi_l, \Pi_e$. A legitimate user has secure connection to the \ac{AP}, if their distance separation is smaller than the distance between that user and any eavesdropper. The number of secure connections to the \ac{AP} accepts an elegant geometric interpretation using the \ac{PVT}: It is equal to the number of legitimate users that fall inside the Voronoi cell $\mathcal{C}_0$ of the point process $\Pi_e \cup \left\{S_0\right\}$~\cite{Pinto2012}, see Fig.~\ref{fig:Eaves} for example illustrations. Conditioned on the area $A$, the \ac{RV} describing the number of secure connections to the \ac{AP}, $N_{\text{in}}$, follows the Poisson distribution ${\text{Po}\!\left(\frac{\lambda_l}{\lambda_e} A\right)}$, where we have to divide by the intensity of the eavesdroppers, $\lambda_e$, to consider the intensity of the \ac{PPP} generating the Voronoi tessellation. The mean and the variance of the \ac{RV} $N_{\text{in}}$ can be expressed in terms of $\mathbb{E}\left\{A\right\}$ and $\mathbb{E}\left\{A^2\right\}$. One has to average the mean and variance of the Poisson  distribution ${\text{Po}\!\left(p A\right)}$ over $A$.
\begin{equation}
\label{eq:MeanVarNin}
\begin{array}{ccl}
\mathbb{E}\left\{N_{{\text{in}}}\right\} &\!\!\!=\!\!\!& p \, \mathbb{E}\left\{A \right\}, \, p\!=\!\frac{\lambda_l}{\lambda_e}.\\
\mathbb{V}{\text{ar}}\left\{N_{{\text{in}}}\right\} &\!\!\!=\!\!\!& p \, \mathbb{E}\left\{A \right\} + p^2 \, \mathbb{E}\left\{A^2 \right\} -p^2\mathbb{E}\left\{A \right\}^2.
\end{array}
\end{equation}
\begin{lemma} The probability that no legetimate user has secure connection to the \ac{AP} can be approximated by $\frac{1}{\left(1+p\,\nu\right)^k}$, where $k,\nu$ are parameters of the Gamma distribution shown in  Table~\ref{table:Table1}. 
\begin{proof}
The \ac{PMF} of the \ac{RV} $N_{\text{in}}$, $f_{N_{\text{in}}}\!\left(n\right)$, is approximated by averaging the Poisson \ac{PDF} ${\text{Po}\!\left(p A\right)}$ over the Gamma approximation for the \ac{PDF} of the cell area. 
\[
f_{N_{\text{in}}}\!\left(n\right) \approx \displaystyle \int\limits_0^\infty \frac{\left(p\, A\right)^n e^{-p\, A}}{n!} \, \frac{ A^{k-1} e^{-A/ \nu}}{\nu^k \Gamma\left(k\right)} {\rm d}A \!=\! \frac{\left( p\,\nu \right)^n \, \Gamma\left(k+n\right) }{ n!\, \Gamma\left(k\right) \left(1+p\,\nu\right)^{n+k}}.
\]
After setting $n\!=\!0$ in the above approximation we get the desired result. 
\end{proof}
\end{lemma}

The approximation for the \ac{CDF} of the \ac{RV} $N_{in}$ can be expressed in terms of the Gaussian hypergeometric function ${}_2F_1$~\cite[pp.~556]{Abramowitz1972}
\begin{equation}
\label{eq:2DInDegreeCDF}
F_{N_{\text{in}}}\!\left(n\right) \approx \displaystyle 1 -  \frac{\left(p\,\nu\right)^{1+n} \Gamma\left(1\!+\!k\!+\!n\right) {}_2F_1\left(1,1\!+\!k\!+\!n,2\!+\!n;\frac{p\,\nu}{1+p\,\nu}\right)}{\left(1\!+\!p\,\nu\right)^{1+k+n} \Gamma\left(k\right) \Gamma\left(2\!+\!n\right)}. 
\end{equation} 

\begin{figure}
\centering
  \includegraphics[width=3.0in]{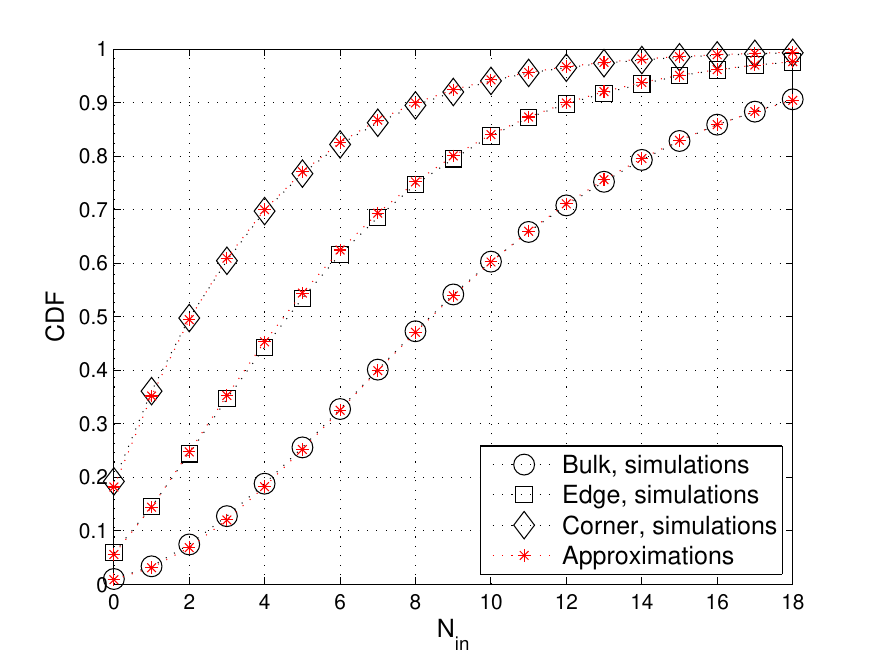}
 \caption{The simulated \ac{CDF} of the \ac{RV} $N_{\text{in}}$ at different locations $S_0$ for the \ac{AP}. The intensities for the legitimate users and the eavesdroppers are $\lambda_l\!=\!10$ and $\lambda_e\!=\!1$. In the simulations, we consider a square with side $L\!=\! 10$ and we place $S_0$ at the corner $\left(0,0\right)$, at the edge $\left(L/2,0\right)$, and in the middle of the square $\left(L/2,L/2\right)$. $20\, 000$ simulations per location. In the approximations, we have used~\eqref{eq:2DInDegreeCDF} with parameters $k,\nu$ available in Table~\ref{table:Table1} for the different locations of $S_0$.}
 \label{fig:InDegreeCDF}
\end{figure}

In Fig.~\ref{fig:InDegreeCDF} we illustrate the approximation accuracy of~\eqref{eq:2DInDegreeCDF} at the corner, at the edge and in the bulk, with parameters $k,\nu$ available in Table~\ref{table:Table1}.  We see very good agreement with the simulated \acp{CDF}.
Let us denote by $N_{\text{out}}$ the \ac{RV} describing the number of legitimate users where the \ac{AP} can securely transmit to. Following the same distance-based rule, the \ac{AP} can securely transmit to a legitimate user if their distance is smaller than the distance between the \ac{AP} and any eavesdropper. The distribution of the \ac{RV} $N_{\text{out}}$ is independent of the location $S_0$. The quantity $\frac{\lambda_l}{\lambda_e+\lambda_l}\!=\! \frac{p}{1+p}$ is the probability that the next user we meet as we move away from the \ac{AP} is legitimate. Therefore $N_{\text{out}}\!=\! n$, if we succeed in meeting $n$ legitimate users before the first eavesdropper. Hence, the \ac{PMF} of the \ac{RV} $N_{\text{out}}$ is Geometric with parameter $\frac{p}{1+p}$~\cite{Pinto2012, Haenggi2009}.  
\begin{equation}
\label{eq:InDegreeCase1}
\begin{array}{ccccc}
f_{N_{\text{out}}}\!\left(n\right) &=& \left(\frac{p}{1+p}\right)^n \frac{1}{1+p}, \,\,\, F_{N_{\text{out}}}\!\left(n\right) &=& 1 - \left(\frac{p}{1+p}\right)^{1+n}, \,\, n\geq 0.
\end{array}
\end{equation}

From~\eqref{eq:InDegreeCase1} we get $\mathbb{E}\left\{N_{{\text{out}}}\right\} = p$ and $\mathbb{V}{\text{ar}}\left\{N_{{\text{out}}}\right\}=p\left(1+p\right)$. The probability that the \ac{AP} has no secure connection to any legitimate user is $f_{N_{\text{out}}}\!\left(0\right)\!=\! \frac{1}{1+p}$. 
\begin{lemma} In regions with boundaries, the mean degrees $\mathbb{E}\left\{N_{{\text{in}}}\right\}$ and $\mathbb{E}\left\{N_{{\text{out}}}\right\}$ are not equal.
\begin{proof}
Since  $\mathbb{E}\left\{A\right\}<1$ along the boundary of a quadrant, see Lemma~\ref{lem:2}, $\mathbb{E}\left\{N_{{\text{out}}}\right\}\!=\!p\!>\!p\,\mathbb{E}\left\{A\right\}\!\stackrel{(a)}{=}\!\mathbb{E}\left\{N_{{\text{in}}}\right\}$, where $(a)$ follows from equation~\eqref{eq:MeanVarNin}. On the other hand, close to the boundary and far from the corner, we may have $\mathbb{E}\left\{A\right\}>1$, see Lemma~\ref{lem:3}, thus $\mathbb{E}\left\{N_{{\text{out}}}\right\}<\mathbb{E}\left\{N_{{\text{in}}}\right\}$. Finally, in the bulk, $\mathbb{E}\left\{A\right\}\!=\!1$, and $\mathbb{E}\left\{N_{{\text{out}}}\right\}=\mathbb{E}\left\{N_{{\text{in}}}\right\}=p$. 
\end{proof}
\end{lemma}

\begin{figure}
\centering
\includegraphics[width=3.0in]{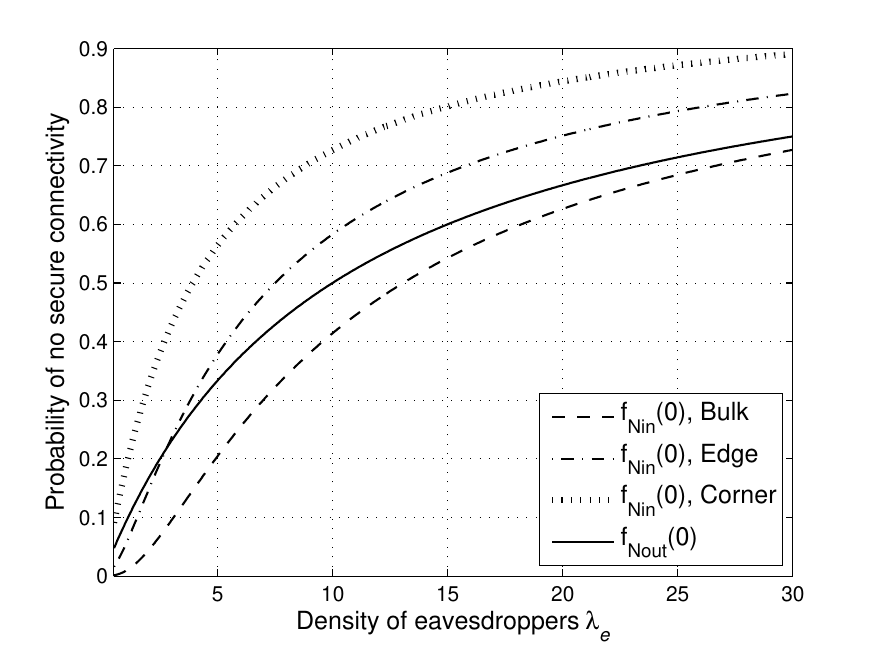}
\caption{Probability there is no secure connection w.r.t. to the intensity of eavesdroppers, while the intensity of legitimate users is $\lambda_l\!=\!10$. The probability there is no secure connection to the \ac{AP}, $f_{N_{in}}\!\left(0\right)$, is approximated by $\frac{1}{\left(1+p\,\nu\right)^k}$ and from the \ac{AP}, $f_{N_{out}}\!\left(0\right)$,  is equal to $\frac{1}{1+p}$, where $p\!=\!\frac{\lambda_l}{\lambda_e}$.}
\label{fig:IsolationProb}
\end{figure}

The relation between the probabilities that no legitimate user has secure connection towards the \ac{AP}, $f_{N_{\text{in}}}\!\left(0\right)$, and from the \ac{AP}, $f_{N_{\text{out}}}\!\left(0\right)$, depends on the location of the \ac{AP}, $S_0$, and the intensity ratio $p$. Verifying the results of~\cite{Pinto2012} for the bulk, it is more probable to have at least one secure connection to the \ac{AP} than from the \ac{AP}, $f_{N_{\text{in}}}\!\left(0\right)\!<\!f_{N_{\text{out}}}\!\left(0\right)$, see Fig.~\ref{fig:IsolationProb}. This is because, a single eavesdropper close to the \ac{AP}, hinders secure connections from the \ac{AP} to the legitimate users, while secure connections towards the \ac{AP} are still possible, e.g., users may be located close to the \ac{AP} but still further from the eavesdroppers. However, this does not hold true in general. In Fig.~\ref{fig:IsolationProb}, we see that at the corner of a quadrant and at the edge of the half-plane, it is more likely to have secure connection from the \ac{AP} to the legitimate users than from the users to the \ac{AP} (for a moderate to high intensity of eavesdroppers). Close to the boundaries, the probability to have secure connections to the \ac{AP} decreases, because the available locations for the legitimate users over there are much less as compared to the bulk. 

\section{Conclusions}
\label{sec:Conclusions}
Instead of running extensive simulations, we have used a low-complexity numerical method for computing the mean cell area in a   homogeneous Poisson Voronoi tessellation for seeds located along and/or close to the boundary of a quadrant. We have shown that the mean cell area falling inside the deployment domain is location-dependent. In addition, we have calculated the second moment of the cell area for a seed at the corner of the quadrant and at the boundary of the half-plane. Even though the distribution of the cell area is not exactly Gamma, the two-parameter Gamma distribution with fitted mean and variance using the method of moments still provides a reasonably good approximation, which might be useful to certain applications. Besides the connectivity of wireless sensor networks with physical layer security, the fitted Gamma distribution can also be used in the performance analysis of finite area cellular networks, e.g., modeling the network load located close to the network borders. In the foreseen deployments of indoor low-power wireless networks, the impact of boundaries cannot be neglected. Another direction for future work is the approximation for the area distribution of edge Voronoi cells due to non-homogeneous \acp{PPP}, i.e., the cells separating regions of different intensities. This may complement existing heuristics for edge detection in high-energy particle physics, see~\cite{Debnath2015, Debnath2016}.   

\section*{Acknowledgement}
This work was supported by the EPSRC grant number EP/N002458/1 for the project Spatially Embedded Networks. All underlying data are provided in full within this paper. The authors would like to thank the anonymous reviewer for providing very constructive feedback.

\end{document}